\documentclass[aps,prc,superscriptaddress,twoside,twocolumn,nofootinbib,10pt,%
showpacs,floatfix]{revtex4-1}

\usepackage{amsmath,amssymb}
\usepackage{graphicx,bm}
\usepackage{slashed}
\usepackage{epstopdf}
\usepackage{ulem} 
\usepackage[usenames]{color}
\usepackage{float}

\begin{document}
\title{ 
Dibaryons with two strange quarks and total spin zero in a constituent quark model 
}

\author{Woosung Park}\email{diracdelta@hanmail.net}\affiliation{Department of Physics and Institute of Physics and Applied Physics, Yonsei University, Seoul 120-749, Korea}
\author{Aaron Park}
\email{lacid0220@naver.com}\affiliation{Department of Physics and Institute of Physics and Applied Physics, Yonsei University, Seoul 120-749, Korea}
\author{Su Houng Lee}\email{suhoung@yonsei.ac.kr}\affiliation{Department of Physics and Institute of Physics and Applied Physics, Yonsei University, Seoul 120-749, Korea}
\date{\today}
\begin{abstract}
We investigate the symmetry property and construct the wave function of  the dibaryon states containing two strange quarks with S=0 in  both the flavor SU(3) symmetric and breaking cases.  
We discuss how  the color $\otimes$ isospin $\otimes$ spin states of dibaryon in the symmetry broking case of  flavor SU(3) can be extracted from the fully antisymmetric states in flavor SU(3).  The stability of the dibaryon against the strong decay into two baryons are then discussed, by using the variational method within a constituent quark model with a confining and color-spin interactions.   To compare our results with that from lattice QCD in flavor SU(3) limit, we search for the stable H-dibaryon in a wide range of $\pi$ meson mass.  We find that with the given potential, there is no compact six quark dibaryon state in the SU(3) flavor symmetry broken case with realistic quark masses as well as in flavor SU(3) symmetric case in a wide range of quark masses.     
\end{abstract}


\maketitle

\section{Introduction}

Since Jaffe~\cite{Jaffe:1976ih,Jaffe:1976ig,Jaffe:1976yi} suggested the possible existence of  tetraquarks and dibrayons in QCD, based on the one-gluon exchange color spin interaction,  multi-quark systems have been explored in various models and searched for experimentally over several decades.  Attempts to find a stable multi-quark systems have been made by many researchers using the chromomagnetic model based on the color spin interaction. For example, for the X(3872), which by now is widely believed to be a $J^{PC}$=$1^{++}$ state, Hogassen~\cite{Hogaasen:2013nca,Hogaasen:2005jv} suggested that it could be a tetraquark state with a strong mixing of the color  octet-octet component of the two quark-antiquark pair within the color spin interaction.  
Silvestre-Brac systematically classified dibaryons consisting of light quarks within the flavor SU(3) symmetry~\cite{SilvestreBrac:1992yg} and those containing two different types of heavy quarks~\cite{Leandri:1995zm}:  These 
papers have discussed the stability of the multi-quark system and their relation to the hyperfine splitting. 
  
The dynamical problem of studying the stability of  the multi-quark system has been studied  mainly using the  variational method with a nonrelativistic Hamiltonian, including the confinement and hyperfine potential.  The ground states of $qq\bar{q}\bar{q}$ systems with L=0, have been extensively calculated with the harmonic oscillator bases in Ref~\cite{SilvestreBrac:1993ss,SilvestreBrac:1993ry}. Calculations based on the simple Gaussian spatial function have been made in Refs.~\cite{Brink:1994ic} and ~\cite{Brink:1998as} to study the stability of $qq\bar{q}\bar{q}$ and $qq\bar{Q}\bar{Q}$ (Q= $c$ or $b$) system respectively.  Two of us~\cite{Park:2013fda} have also introduced the correlation between quarks in the Gaussian spatial function to investigate its effect on the stability of $qq\bar{Q}\bar{Q}$ (Q= $c$ or $b$) system. In addition to these variational methods,  
a powerful tool, using the hyperspherical harmonic basis functions, have been developed in Ref.~\cite{Vijande:2007rf,Vijande:2007fc} to solve the four-body problem in  the tetraquark.

The stability study in the dibaryon sector has also been pursed with  several other  models\cite{Gal:2015rev,Sakai:1999qm,Kodama:1994np,Oka:1990vx}.   For dibaryons containing light quarks only, the H-dibaryon is expected to be the most stable state as the color spin hyperfine splitting is most attractive,   even compared to the two $\Lambda$ baryons. 
While the model study based on the Goldstone exchange interaction cast some doubt on  the existence of the H-dibaryon~\cite{Stancu:1998ca}, recent results from lattice QCD have suggested that the H-dibaryon would be
bound for massive $\pi$ mass~\cite{Beane:2010hg,Inoue:2010es}. Also, the H-dibaryon was found to be bound in a chiral constituent quark model calculation~\cite{Carames:2012zz}.

The purpose of this paper is to find the  color $\otimes$ isospin $\otimes$ spin states in the SU(3) flavor symmetry broken case appropriate for the dibaryon containing two strange quarks with S=0 that is compatible with a symmetric spatial wave function, and then calculate the mass of the dibaryon by using the variational method, with the color-spin hyperfine potential introduced in Ref.~\cite{Bhaduri:1981pn}. In particular, by going through  the systematic construction of the color $\otimes$ isospin $\otimes$ spin states in general, we find the corresponding state for the symmetry broken case of flavor SU(3). Lastly, we search for the stability of the H-dibaryon in the flavor symmetric limit of SU(3) as a function of the pion mass.  Through this work, we will be able to verify if a compact H dibaryon exists in the symmetry breaking case or the symmetric limit of SU(3) within the given Hamiltonian.  If the recent lattice result of stable H dibaryon in the massive pion case turns out to be valid, our result strongly suggests that the state should be a hadronic bound state and/or loosely bound molecular state.     

This paper is organized as follows. We first present the Hamiltonian, and calculate the mass of both baryon octet and decuplet to determine the fitting parameters of the model, and construct the spatial wave function of the dibaryon in section II. In section III, we construct the color $\otimes$ isospin $\otimes$ spin states in both the flavor SU(3) symmetric and broken cases, establishing the relation between the two cases.  We show the numerical results obtained from the variational method in section IV. Finally, we summarize the results in section V. The appendices includes some details of the calculations.

\section{Hamiltonian}

To investigate the stability of dibaryon, we adopt the following nonrelativistic Hamiltonian that  contains the confinement and hyperfine potential for the color and spin interaction. 
\begin{align}
H=\sum_{i=1}^{6}(m_{i}+\frac{\textbf{p}^2_i}{2m_i})-\frac{3}{16}\sum_{i<j}^{6}
\lambda^c_i\lambda^c_j(V^{C}_{ij}+V^{SS}_{ij}),
\label{eq-hamiltonian}
\end{align}
where $m_i$'s are the quark masses, $\lambda^c_i/2$ the color operator of the $i$'th quark for the color SU(3), and $V^{C}_{ij}$ and $V^{SS}_{ij}$ the confinement and hyperfine potential, respectively.  For the  confinement potential, we take half-linearising potential and Coulomb potential as follows:
\begin{align}
V^{C}_{ij}=-\frac{\kappa}{r_{ij}}+\frac{(r_{ij})^{1/2}}{a_0}-D. 
\label{eq-confinement}
\end{align}
Here, the first term comes from the perturbative one-gluon exchange, and the second confining part will more likely give stability to the dibaryon and other multi-quark system compared to taking a linearly rising potential but will not change the main result. 

For the hyperfine potential, we take the potential to be dependent upon spin interaction as  follows:
\begin{align}
V^{SS}_{ij}=\frac{1}{m_im_jc^4}\frac{\hbar^2c^2{\kappa}^{\prime}}{(r_{0ij})}\frac{e^{-(r_{ij})^2/(r_{0ij})^2}}{r_{ij}}
{\sigma}_i\cdot{\sigma}_j. 
\label{eq-hyperfine}
\end{align}
Here, $r_{ij}$ is the distance between interquarks, $\mid\textbf{r}_i-\textbf{r}_j\mid$, and both $(r_{0ij})$ and ${\kappa}^{\prime}$ are chosen to depend on the masses of interquarks, given by,
\begin{align}
&r_{0ij}=1/(\alpha+\beta \frac{m_im_j}{m_i+m_j}), \nonumber \\
&{\kappa}^{\prime}={\kappa}_0(1+\gamma \frac{m_im_j}{m_i+m_j}). \label{eq-hyperfine2}
\end{align}
The hyperfine potential splits the baryon octet as well as baryon decuplet masses. Moreover, in the heavy quark mass limit $m_i \to\infty$, the functional form of the hyperfine potential in Eq.~(\ref{eq-hyperfine}) approaches  $1/(m_im_j)$ $\delta(r)$. By introducing the additional parameters in   Eq.~(\ref{eq-hyperfine2}), one is able to reproduce the experimental observation that the mass differences between psudoscalar and vector meson for $q\bar{q}$ states decrease slower as a function of the masses than that given by the inverse mass relation  given in Eq.~(\ref{eq-hyperfine}). 

In the Hamiltonian, the fitting parameters have been chosen to reproduce the experimental values of both the baryon octet and decuplet masses using the  variational method and typically used constituent quark masses. The fitting parameters are given in Table~\ref{fitting-parameter-1}.
\begin{table}[htdp]
\caption{ Parameters fitted to the experimental  baryon octet and
decuplet masses. The third row indicates the units of the parameters using $\hbar$ = $c$ =1.}
\begin{center}
\begin{tabular}{c|c|c|c|c|c|c|c|c}
\hline \hline
  $\gamma$  & $\kappa$  & $a_0$  & D & ${\kappa}_0$ & $\alpha$ &$\beta$ & $m_u$  & $m_s$     \\
\hline
   0.309     &  0.123  & 1.049  & 0.994  & 0.35    &  2.105   & 9.164        & 0.347  & 0.596  \\
\hline 
  $\rm{(GeV)^{-1}}$    &      & $\rm{(GeV)^{-3/2}}$   &   $\rm{GeV}$  & $\rm{GeV}$ & $\rm{GeV}$ &                 & $\rm{GeV}$   & $\rm{GeV}$      \\
\hline \hline  
\end{tabular}
\end{center}
\label{fitting-parameter-1}
\end{table}

As in our previous work~\cite{Park:2015nha},  we will choose the color $\otimes$ isospin $\otimes$ spin state and the spatial wave function with a simple Gaussian form for  the baryons and calculate the mass of both baryon octet and decuplet with the new fitting parameters using the variational method. The masses are given in Table~\ref{fitting-mass-1}. 
\begin{table}[htdp]
\caption{Tthe mass of baryon octet and decuplet obtained from the variational method. The third row shows the experimental data~\cite{Agashe:2014kda}(unit: $\rm{GeV}$). }
\begin{center}
\begin{tabular}{c|c|c|c|c|c|c|c|c}
\hline \hline
(I,S)      & ($\frac{1}{2}$,$\frac{1}{2}$) &($\frac{1}{2}$,$\frac{1}{2}$)& (0,$\frac{1}{2}$) & (1,$\frac{1}{2}$) &($\frac{1}{2}$,$\frac{3}{2}$) &(1,$\frac{3}{2}$) &($\frac{3}{2}$,$\frac{3}{2}$)  & (0,$\frac{3}{2}$)  \\
           & N, P     & $\Xi$    & $\Lambda$   & $\Sigma$     & $\Xi^*$ & $\Sigma^*$  &  $\Delta$ &$\Omega$   \\
\hline
Mass  & 0.974   & 1.344   & 1.115        & 1.217    & 1.554       & 1.398   & 1.233 &1.7  \\                     
\hline
Exp       &  0.938   &  1.314    &              & 1.189    & 1.53     & 1.382       & 1.23       & \\
            & $\sim$  & $\sim$     & 1.115  & $\sim$    &  $\sim$ & $\sim$    & $\sim$ &1.672 \\
          &  0.939   &  1.321        &              & 1.197     & 1.531  & 1.387      & 1.234     &  \\  
\hline \hline
\end{tabular}
\end{center}
\label{fitting-mass-1}
\end{table}

In calculating the mass of dibaryon containing two strange quarks with spin=0, we need the spatial function appropriate for six-quark system with a certain symmetry property. Since we will construct the color $\otimes$ isospin $\otimes$ spin state of the dibaryon to be antisymmetric among particles 1, 2, 3, and 4, and at the same time antisymmetric between particles 5 and 6, which will be denoted by $\{1234\}\{56\}$, the symmetry property of spatial function should be symmetric among particles 1, 2, 3, and 4, and at the same time symmetric between particles 5 and 6, due to Pauli principle; we will denote the symmetry property of the spatial function by [1234][56].
 
In order to describe the six-quark system, we consider the center of mass frame, so that the number of suitable Jacobian coordinates of the system is reduced to 5. The five Jacobian coordinates are given by
\begin{align}
&\pmb{x_1}=\frac{1}{\sqrt{2}}(\textbf{r}_5-\textbf{r}_6),\quad
\pmb{x_2}=\frac{1}{2}(\textbf{r}_1-\textbf{r}_2+\textbf{r}_3-\textbf{r}_4),\nonumber\\
&\pmb{x_3}=\frac{1}{2}(\textbf{r}_1-\textbf{r}_2-\textbf{r}_3+\textbf{r}_4),\quad
\pmb{x_4}=\frac{1}{2}(\textbf{r}_1+\textbf{r}_2-\textbf{r}_3-\textbf{r}_4),\nonumber\\
&\pmb{x_5}=\frac{1}{\sqrt{12}}(\textbf{r}_1+\textbf{r}_2+\textbf{r}_3+\textbf{r}_4-2\textbf{r}_5-2\textbf{r}_6).
\label{eq-jacobian}
\end{align}
Then, we can construct the spatial wave function of the dibaryon in a Gaussian form, which will be used to carry out
the variational method, given by, 
\begin{align}
R=\exp[-&(a(\pmb{x_1})^2+b(\pmb{x_2})^2+b(\pmb{x_3})^2+b(\pmb{x_4})^2+\nonumber\\
&c(\pmb{x_5})^2)],
\label{eq-spatial}
\end{align}
where $a$, $b$, and $c$ are variational parameters. It is easily found that the symmetry of the spatial wave function 
has the [1234][56] property, required by the color $\otimes$ isospin $\otimes$ spin state of the dibaryon.

\section{Classification of Dibaryon containing two strange quarks with spin=0}

\subsection{ The state of Dibaryon with respect to isospin states}

In this section, we investigate the state of the dibaryon containing two strange quarks with S=0, with flavor symmetry represented by SU(2). The symmetry breaking of SU(3) in flavor part is caused by taking the strange quark mass to be heavier compared to $m_u$ (=$m_d$). When we choose the
symmetry of spatial function to be symmetric under the exchange of any two particles among 1, 2, 3, and 4, and at the same time symmetric under the exchange of two particles between 5 and 6, the fixing of the position of two strange quarks onto the fifth and sixth is convenient to classify the dibaryon state. The four light quarks except for two strange quarks would be characterized by introducing Young tableau corresponding to isospin states, as follows;
\begin{align}
\begin{tabular}{c}
$ I^0$ ;
\end{tabular}
\begin{tabular}{|c|c|}
\hline
$\quad$ &$\quad$    \\
\cline{1-2}
$\quad$ & $\quad$   \\  
\hline  
\end{tabular}_{\quad [1]}
\quad
\begin{tabular}{c}
$ I^1$ ;
\end{tabular}
\begin{tabular}{|c|c|c|}
\hline
$\quad$ &$\quad$ & $\quad$  \\
\cline{1-3}
\multicolumn{1}{|c|}{$\quad$}    \\  
\cline{1-1}  
\end{tabular}_{ [3]}
 \quad
\begin{tabular}{c}
$I^2$ ;
\end{tabular}
\begin{tabular}{|c|c|c|c|}
\hline
$\quad$ &$\quad$ & $\quad$ & $\quad$   \\
\cline{1-4}  
\end{tabular}_{\quad [5]}
\label{eq-I}
\end{align}
Here, the dimension of each isospin state is shown below the Young tableau.

In our case where we choose the symmetry of spatial function of the dibaryon to be [1234][56], the color $\otimes$ isospin $\otimes$ spin state of the dibaryon will be chosen to be $\{1234\}\{56\}$ in order to satisfy Pauli principle. When we fix the positions of the two strange quarks onto the fifth and sixth, the partly antisymmetric state of the color $\otimes$ isospin $\otimes$ spin state can be easily obtained from classifying the multiplets of the direct four product of $[12]_{CIS}$ multiplied by direct two product of $[6]_{CS}$, which represent the fundamental representation of  $SU(12)_{CIS}$ and  $SU(6)_{CS}$, respectively. Since the state of $\{1234\}$ by the direct four product of $[12]_{CIS}$ gives the multiplet with dimension 495 corresponding to Young tableau $[1^4]$, and the state of $\{56\}$ by the direct two product of $[6]_{CS}$ the multiplet with dimension 15 corresponding to Young tableau $[1^2]$, the dimension of $\{1234\}\{56\}$ is $495\times15 = 7425$. This $\{1234\}\{56\}$ state can be decomposed into the direct sum of representation ($[2]_{I}$, $[6]_{CS}$). The $\{1234\}\{56\}$ state with dimension 7425 decomposed into the direct sum of representation ($[2]_{I}$, $[6]_{CS}$) is given in Eq.~(\ref{eq-CIS}). By using the Young tableau, we can easily find that the multiplets in Eq.~(\ref{eq-CIS}) is $\{1234\}\{56\}$. Because the color state of the dibaryon is supposed to be a physically observable color singlet which corresponds to Young tableau [2,2,2], and the spin state of the dibaryon in our works is confined to S=0 which corresponds to Young tableau [3,3], we need the possible color $\otimes$ spin states by combining color singlet with S=0 state, which will be called CS coupling scheme. Then, only the color $\otimes$ spin states obtained from the CS coupling scheme are allowed among the multiplets in Eq.~(\ref{eq-CIS}). Since Young tableau corresponding to the color $\otimes$ spin states are [3,3], [2,2,1,1], and $[1^6]$, the $\{1234\}\{56\}$ states of the dibaryon with respect to S=0 are given as, $([1]_I, [490]_{CS})$, $([1]_I, [189]_{CS})$, $([3]_I, [189]_{CS})$, $([5]_I, [189]_{CS})$, and $([5]_I, [1]_{CS})$.   We note that another Young tableau [4,1,1] obtained from the CS coupling scheme is excluded, because the Young tableau [4,1,1] does not belong to the multiplets of   the color $\otimes$ spin state in Eq.~(\ref{eq-CIS}).
\begin{widetext}\allowdisplaybreaks

\begin{align}
&\{1234\}\{56\}_{[7425]}= \nonumber \\
&\begin{tabular}{|c|c|}
\hline
$\quad$  & $\quad$   \\
\cline{1-2}
$\quad$  &  $\quad$   \\
\cline{1-2}
\end{tabular}_{\quad [1]}
\otimes
\begin{tabular}{|c|c|c|}
\hline
           $\quad$ &  $\quad$  & $\quad$    \\
\cline{1-3} $\quad$  & $\quad$ & $\quad$   \\
\hline
\end{tabular}_{\quad [490]}
\oplus 
\begin{tabular}{|c|c|}
\hline
$\quad$  & $\quad$   \\
\cline{1-2}
$\quad$  &  $\quad$   \\
\cline{1-2}
\end{tabular}_{\quad [1]}
\otimes
\begin{tabular}{|c|c|}
\hline
$\quad$ & $\quad$   \\
\cline{1-2}
$\quad$ &$\quad$    \\
\cline{1-2}
\multicolumn{1}{|c|}{$\quad$}  \\
\cline{1-1}
\multicolumn{1}{|c|}{$\quad$}  \\
\cline{1-1}
\end{tabular}_{ [189]} 
\oplus 
\begin{tabular}{|c|c|}
\hline
$\quad$  & $\quad$   \\
\cline{1-2}
$\quad$  &  $\quad$   \\
\cline{1-2}
\end{tabular}_{\quad [1]}
\otimes
\begin{tabular}{|c|c|c|}
\hline
           $\quad$ &  $\quad$  & $\quad$    \\
\cline{1-3} 
\multicolumn{1}{|c|}{$\quad$}&\multicolumn{1}{c|}{$\quad$}  \\
\cline{1-2}
\multicolumn{1}{|c|}{$\quad$}  \\
\cline{1-1}
\end{tabular}_{ [896]}
\oplus  
\begin{tabular}{|c|c|c|}
\hline
           $\quad$ &  $\quad$  & $\quad$    \\
\cline{1-3} 
\multicolumn{1}{|c|}{$\quad$} \\
\cline{1-1}
\end{tabular}_{ [3]}
\otimes
\begin{tabular}{|c|c|}
\hline
$\quad$ & $\quad$   \\
\cline{1-2}
$\quad$ &$\quad$    \\
\cline{1-2}
\multicolumn{1}{|c|}{$\quad$}  \\
\cline{1-1}
\multicolumn{1}{|c|}{$\quad$}  \\
\cline{1-1}
\end{tabular}_{ [189]} 
\oplus  \nonumber \\
&\begin{tabular}{|c|c|c|}
\hline
           $\quad$ &  $\quad$  & $\quad$    \\
\cline{1-3} 
\multicolumn{1}{|c|}{$\quad$} \\
\cline{1-1}
\end{tabular}_{ [3]}
\otimes
\begin{tabular}{|c|c|c|}
\hline
           $\quad$ &  $\quad$  & $\quad$    \\
\cline{1-3} 
\multicolumn{1}{|c|}{$\quad$}&\multicolumn{1}{c|}{$\quad$}  \\
\cline{1-2}
\multicolumn{1}{|c|}{$\quad$}  \\
\cline{1-1}
\end{tabular}_{ [896]}
\oplus  
\begin{tabular}{|c|c|c|}
\hline
           $\quad$ &  $\quad$  & $\quad$    \\
\cline{1-3} 
\multicolumn{1}{|c|}{$\quad$} \\
\cline{1-1}
\end{tabular}_{ [3]}
\otimes
\begin{tabular}{|c|c|c|}
\hline
           $\quad$ &  $\quad$  & $\quad$    \\
\cline{1-3} 
\multicolumn{1}{|c|}{$\quad$} \\
\cline{1-1}
\multicolumn{1}{|c|}{$\quad$}  \\
\cline{1-1}
\multicolumn{1}{|c|}{$\quad$}  \\
\cline{1-1}
\end{tabular}_{ [280]}
\oplus 
\begin{tabular}{|c|c|c|}
\hline
           $\quad$ &  $\quad$  & $\quad$    \\
\cline{1-3} 
\multicolumn{1}{|c|}{$\quad$} \\
\cline{1-1}
\end{tabular}_{ [3]}
\otimes
\begin{tabular}{|c|c|}
\hline
           $\quad$ &  $\quad$    \\
\cline{1-2} 
\multicolumn{1}{|c|}{$\quad$} \\
\cline{1-1}
\multicolumn{1}{|c|}{$\quad$}  \\
\cline{1-1}
\multicolumn{1}{|c|}{$\quad$}  \\
\cline{1-1}
\multicolumn{1}{|c|}{$\quad$}  \\
\cline{1-1}
\end{tabular}_{ [35]}
\oplus
\begin{tabular}{|c|c|c|}
\hline
           $\quad$ &  $\quad$  & $\quad$    \\
\cline{1-3} 
\multicolumn{1}{|c|}{$\quad$} \\
\cline{1-1}
\end{tabular}_{ [3]}
\otimes
\begin{tabular}{|c|c|}
\hline
           $\quad$ &  $\quad$    \\
\cline{1-2} 
$\quad$ & $\quad$   \\
\cline{1-2}
$\quad$ & $\quad$   \\
\cline{1-2}
\end{tabular}_{\quad [175]}
\oplus  \nonumber \\
&\begin{tabular}{|c|c|c|c|}
\hline
           $\quad$ &  $\quad$  & $\quad$ &  $\quad$   \\
\cline{1-4} 
\end{tabular}_{\quad [5]}
\otimes
\begin{tabular}{|c|}
\hline
           $\quad$    \\
\cline{1-1} 
   $\quad$    \\
\cline{1-1} 
   $\quad$    \\
\cline{1-1} 
   $\quad$    \\
\cline{1-1} 
   $\quad$    \\
\cline{1-1} 
   $\quad$    \\
\cline{1-1} 
\end{tabular}_{\quad [1]}
\oplus 
\begin{tabular}{|c|c|c|c|}
\hline
           $\quad$ &  $\quad$  & $\quad$ &  $\quad$   \\
\cline{1-4} 
\end{tabular}_{\quad [5]}
\otimes
\begin{tabular}{|c|c|}
\hline
$\quad$ & $\quad$   \\
\cline{1-2}
$\quad$ &$\quad$    \\
\cline{1-2}
\multicolumn{1}{|c|}{$\quad$}  \\
\cline{1-1}
\multicolumn{1}{|c|}{$\quad$}  \\
\cline{1-1}
\end{tabular}_{ [189]} 
\oplus 
\begin{tabular}{|c|c|c|c|}
\hline
           $\quad$ &  $\quad$  & $\quad$ &  $\quad$   \\
\cline{1-4} 
\end{tabular}_{\quad [5]}
\otimes
\begin{tabular}{|c|c|}
\hline
           $\quad$ &  $\quad$    \\
\cline{1-2} 
\multicolumn{1}{|c|}{$\quad$} \\
\cline{1-1}
\multicolumn{1}{|c|}{$\quad$}  \\
\cline{1-1}
\multicolumn{1}{|c|}{$\quad$}  \\
\cline{1-1}
\multicolumn{1}{|c|}{$\quad$}  \\
\cline{1-1}
\end{tabular}_{ [35]}
\label{eq-CIS}
\end{align}

\end{widetext}
\subsection{ Isospin $\otimes$ color $\otimes$ spin state of the dibaryon}
Before constructing the color $\otimes$ isospin $\otimes$ spin state, we emphasize that the $\{1234\}\{56\}$ state of the dibaryon containing two strange quarks with S=0 can be obtained from the CS coupling scheme with respect to isospin states, but also be derived from a fully antisymmetric color $\otimes$ flavor $\otimes$ spin state, in which flavor is SU(3). For this reason, we will specifically indicate the flavor state in terms of SU(3) symmetry; as
we will show later, we find that the symmetry breaking of SU(3) from a fully antisymmetric color $\otimes$ flavor $\otimes$ spin state of the dibaryon lead to the $\{1234\}\{56\}$ state.

As was shown in a previous paper by two of us~\cite{Park:2015nha}, the color singlet and S=0 state of the dibaryon are given by the corresponding Young-Yamanouchi basis, respectively; 
\begin{itemize}
\item Color singlet : 5 basis functions with Young tableau [2,2,2] 
\begin{align}
&\begin{tabular}{c}
$\vert C_1 \rangle$=
\end{tabular}
\begin{tabular}{|c|c|}
\hline
1 & 2   \\
\cline{1-2}
3 &  4  \\
\cline{1-2}
5 &  6  \\
\hline
\end{tabular}
\quad
\begin{tabular}{c}
$\vert C_2 \rangle$=
\end{tabular}
\begin{tabular}{|c|c|}
\hline
1 & 3   \\
\cline{1-2}
2 &  4  \\
\cline{1-2}
5 &  6  \\
\hline
\end{tabular} 
\quad
\begin{tabular}{c}
$\vert C_3 \rangle$=
\end{tabular}
\begin{tabular}{|c|c|}
\hline
1 & 2   \\
\cline{1-2}
3 &  5  \\
\cline{1-2}
4 &  6  \\
\hline
\end{tabular} 
\quad
\begin{tabular}{c}
$\vert C_4 \rangle$=
\end{tabular}
\begin{tabular}{|c|c|}
\hline
1 & 3   \\
\cline{1-2}
2 &  5  \\
\cline{1-2}
4 &  6  \\
\hline
\end{tabular}
\nonumber \\
&\begin{tabular}{c}
$\vert C_5 \rangle$=
\end{tabular}
\begin{tabular}{|c|c|}
\hline
1 & 4   \\
\cline{1-2}
2 &  5  \\
\cline{1-2}
3 &  6  \\
\hline
\end{tabular} 
\label{eq-color}
\end{align}
\item S=0 : 5 basis functions with Young tableau [3,3] 
\begin{align}
&\begin{tabular}{c}
$\vert S^0_1 \rangle$=
\end{tabular}
\begin{tabular}{|c|c|c|}
\hline
                1   & 2   & 3    \\
\cline{1-3} 4  &  5 & 6  \\
\hline
\end{tabular}
\begin{tabular}{c}
$\vert S^0_2 \rangle$=
\end{tabular}
\begin{tabular}{|c|c|c|}
\hline
                1   & 2   & 4    \\
\cline{1-3} 3  &  5 & 6  \\
\hline
\end{tabular}
\begin{tabular}{c}
$\vert S^0_3 \rangle$=
\end{tabular}
\begin{tabular}{|c|c|c|}
\hline
                1   & 3   & 4    \\
\cline{1-3} 2  &  5 & 6  \\
\hline
\end{tabular}
\nonumber  \\
&\begin{tabular}{c}
$\vert  S^0_4 \rangle$=
\end{tabular}
\begin{tabular}{|c|c|c|}
\hline
                1   & 2   & 5    \\
\cline{1-3} 3  &  4 & 6  \\
\hline
\end{tabular}
\begin{tabular}{c}
$\vert  S^0_5 \rangle$=
\end{tabular}
\begin{tabular}{|c|c|c|}
\hline
                1   & 3   & 5    \\
\cline{1-3} 2  &  4 & 6  \\
\hline
\end{tabular}
\label{eq-spin}
\end{align}
\end{itemize}

In order to construct the $\{1234\}\{56\}$ state of the dibaryon, we need to know the color $\otimes$ spin state, which can be derived from the CS coupling scheme. The CS coupling scheme is technically completed by using  Clebsch-Gordan (CG) coefficient, given by the following formula~\cite{Stancu:1991rc},   
\begin{align}
&S([f^{\prime}]p^{\prime}q^{\prime}y^{\prime}[f^{\prime\prime}]p^{\prime\prime}q^{\prime\prime}y^{\prime\prime}\vert[f]pqy)=\nonumber\\
&K([f^{\prime}]p^{\prime}[f^{\prime\prime}]p^{\prime\prime}\vert[f]p)
S([f^{\prime}_{p^{\prime}}]q^{\prime}y^{\prime}[f^{\prime\prime}_{p^{\prime\prime}}]q^{\prime\prime}y^{\prime\prime}\vert[f_p]qy),
\label{eq-CG}
\end{align}
where S in the left-hand (right-hand) side is a CG coefficients of $S_n$ ($S_{n-1}$) permutation group, and K is an isoscalar factor, which is called K matrix that factorizes the CG coefficients of $S_n$ into a CG coefficients of $S_{n-1}$ multiplied by the isoscalar factor. In this notation,
$[f_p]$, which represent the Young tableau associated to $S_{n-1}$, can be obtained from $[f]$, the Young tableau of $S_n$, where the $pqy$ represents the row positions of the last three particles in the Young tableau $[f]$, by removing the n-th particle.  In order to obtain the CG coefficients of $S_n$, we repeat the process of factorizing the CG coefficients of $S_6$ further, until Eq.~(\ref{eq-CG}) is extended into the following formula~\cite{Pepin:2001is}, 
\begin{align}                                     
&S([f^{\prime}]p^{\prime}q^{\prime}y^{\prime}r^{\prime}[f^{\prime\prime}]p^{\prime\prime}q^{\prime\prime}y^{\prime\prime}r^{\prime\prime}\vert[f]pqyr)=
\nonumber\\
&K([f^{\prime}]p^{\prime}[f^{\prime\prime}]p^{\prime\prime}\vert[f]p) K([f^{\prime}_{p^{\prime}}]q^{\prime}[f^{\prime\prime}_{p^{\prime\prime}}]q^{\prime\prime}\vert[f_p]q) \nonumber\\
&K([f^{\prime}_{p^{\prime}q^{\prime}}]y^{\prime}[f^{\prime\prime}_{p^{\prime\prime}q^{\prime\prime}}]y^{\prime\prime}\vert[f_{pq}]y)
S([f^{\prime}_{p^{\prime}q^{\prime}y^{\prime}}]r^{\prime}[f^{\prime\prime}_{p^{\prime\prime}q^{\prime\prime}y^{\prime\prime}}]r^{\prime\prime}\vert[f_{pqy}]r),
\label{eq-CG-1}
\end{align}
where S in the third row is the CG coefficient of $S_3$. When we find out the CG coefficients using Eq.~(\ref{eq-CG-1}), we use the results obtained by Stancu and Pepin~\cite{Stancu:1999qr} about 
the relevant isoscalar factors for $S_4$, $S_5$, and $S_6$ appearing in Eq.~(\ref{eq-CG-1}), as showed in the previous our paper~\cite{Park:2015nha}.

Then, we can calculate the CG coefficients between Young tableau [2,2,2] of color singlet state and [3,3] of S=0 state in making the representation of [3,3], [2,2,1,1], and $[1^6]$ of the CS coupling scheme. The color $\otimes$ spin state will be denoted by $\vert C,S^0 \rangle$, all of which  are given in Eq.~(\ref{eq-CS-1},\ref{eq-CS-2},\ref{eq-CS-3}). 

We note that the basis function of  Young tableau is expressed by Young-Yamanouchi representation, whose  symmetry property is symmetric with respect to any neighboring particles that lie in the same row, and is antisymmetric with respect to any neighboring particles that lie in the same column. The color $\otimes$ spin state, which consist of the linear sum of combining the color singlet state with S=0 state will be presented in Appendix B. 

The CS coupling scheme, which represent the color $\otimes$ spin states for Young tableaux [3,3], [2,2,1,1], and $[1^6]$ are given as follows;
\begin{itemize}
\item CS coupling with Young tableau [2,2,1,1] : 9 bases functions
\begin{align}
\begin{tabular}{c}
$\vert [C,S^0]_1 \rangle$=
\end{tabular}
\begin{tabular}{|c|c|}
\hline
1 & 2   \\
\cline{1-2}
3 &  4  \\
\cline{1-2}
\multicolumn{1}{|c|}{5}  \\
\cline{1-1}
\multicolumn{1}{|c|}{6}  \\
\cline{1-1}
\end{tabular}
\quad
\begin{tabular}{c}
$\vert [C,S^0]_2 \rangle$=
\end{tabular}
\begin{tabular}{|c|c|}
\hline
1 & 3   \\
\cline{1-2}
2 &  4  \\
\cline{1-2}
\multicolumn{1}{|c|}{5}  \\
\cline{1-1}
\multicolumn{1}{|c|}{6}  \\
\cline{1-1}
\end{tabular} 
\quad
\begin{tabular}{c}
$\vert [C,S^0]_3 \rangle$=
\end{tabular}
\begin{tabular}{|c|c|}
\hline
1 & 2   \\
\cline{1-2}
3 &  5  \\
\cline{1-2}
\multicolumn{1}{|c|}{4}  \\
\cline{1-1}
\multicolumn{1}{|c|}{6}  \\
\cline{1-1}
\end{tabular} 
\nonumber  \\
\begin{tabular}{c}
$\vert [C,S^0]_4 \rangle$=
\end{tabular}
\begin{tabular}{|c|c|}
\hline
1 & 3   \\
\cline{1-2}
2 &  5  \\
\cline{1-2}
\multicolumn{1}{|c|}{4}  \\
\cline{1-1}
\multicolumn{1}{|c|}{6}  \\
\cline{1-1}
\end{tabular}
\quad
\begin{tabular}{c}
$\vert [C,S^0]_5 \rangle$=
\end{tabular}
\begin{tabular}{|c|c|}
\hline
1 & 4   \\
\cline{1-2}
2 &  5  \\
\cline{1-2}
\multicolumn{1}{|c|}{3}  \\
\cline{1-1}
\multicolumn{1}{|c|}{6}  \\
\cline{1-1}
\end{tabular} 
\quad
\begin{tabular}{c}
$\vert [C,S^0]_6 \rangle$=
\end{tabular}
\begin{tabular}{|c|c|}
\hline
1 & 2   \\
\cline{1-2}
3 &  6  \\
\cline{1-2}
\multicolumn{1}{|c|}{4}  \\
\cline{1-1}
\multicolumn{1}{|c|}{5}  \\
\cline{1-1}
\end{tabular}
\nonumber  \\
\begin{tabular}{c}
$\vert [C,S^0]_7 \rangle$=
\end{tabular}
\begin{tabular}{|c|c|}
\hline
1 & 3   \\
\cline{1-2}
2 &  6  \\
\cline{1-2}
\multicolumn{1}{|c|}{4}  \\
\cline{1-1}
\multicolumn{1}{|c|}{5}  \\
\cline{1-1}
\end{tabular} 
\quad
\begin{tabular}{c}
$\vert [C,S^0]_8 \rangle$=
\end{tabular}
\begin{tabular}{|c|c|}
\hline
1 & 4   \\
\cline{1-2}
2 &  6  \\
\cline{1-2}
\multicolumn{1}{|c|}{3}  \\
\cline{1-1}
\multicolumn{1}{|c|}{5}  \\
\cline{1-1}
\end{tabular}
\quad
\begin{tabular}{c}
$\vert [C,S^0]_9 \rangle$=
\end{tabular}
\begin{tabular}{|c|c|}
\hline
1 & 5   \\
\cline{1-2}
2 &  6  \\
\cline{1-2}
\multicolumn{1}{|c|}{3}  \\
\cline{1-1}
\multicolumn{1}{|c|}{4}  \\
\cline{1-1}
\end{tabular}
\label{eq-CS-1}
\end{align}
\\
\item CS coupling with Young tableau [3,3] : 5 bases functions
\begin{align}
&\begin{tabular}{c}
$\vert [C,S^0]_1 \rangle$=
\end{tabular}
\begin{tabular}{|c|c|c|}
\hline
                1   & 2   & 3    \\
\cline{1-3} 4  &  5 & 6  \\
\hline
\end{tabular}
\begin{tabular}{c}
$\vert [C,S^0]_2 \rangle$=
\end{tabular}
\begin{tabular}{|c|c|c|}
\hline
                1   & 2   & 4    \\
\cline{1-3} 3  &  5 & 6  \\
\hline
\end{tabular}
\begin{tabular}{c}
$\vert [C,S^0]_3 \rangle$=
\end{tabular}
\begin{tabular}{|c|c|c|}
\hline
                1   & 3   & 4    \\
\cline{1-3} 2  &  5 & 6  \\
\hline
\end{tabular}
\nonumber  \\
&\begin{tabular}{c}
$\vert  [C,S^0]_4 \rangle$=
\end{tabular}
\begin{tabular}{|c|c|c|}
\hline
                1   & 2   & 5    \\
\cline{1-3} 3  &  4 & 6  \\
\hline
\end{tabular}
\begin{tabular}{c}
$\vert  [C,S^0]_5 \rangle$=
\end{tabular}
\begin{tabular}{|c|c|c|}
\hline
                1   & 3   & 5    \\
\cline{1-3} 2  &  4 & 6  \\
\hline
\end{tabular}
\label{eq-CS-2}
\end{align}
\\
\item CS coupling with Young tableau $[1^6]$ : 1 basis function
\begin{align}
\begin{tabular}{c}
$\vert [C,S^0] \rangle$=
\end{tabular}
&\frac{1}{\sqrt{5}}
\nonumber \\
\Big[
- &\begin{tabular}{|c|c|}
\hline
1 & 2   \\
\cline{1-2}
3 &  4  \\
\cline{1-2}
5 &  6  \\
\hline
\end{tabular}_{\quad C}
\otimes
\begin{tabular}{|c|c|c|}
\hline
                1   & 3   & 5    \\
\cline{1-3} 2  &  4 & 6  \\
\hline
\end{tabular}_{\quad S^0}
+
\begin{tabular}{|c|c|}
\hline
1 & 3  \\
\cline{1-2}
2 &  4  \\
\cline{1-2}
5 &  6  \\
\hline
\end{tabular}_{\quad C}
\otimes
\begin{tabular}{|c|c|c|}
\hline
                1   & 2   & 5    \\
\cline{1-3}  3 &  4 & 6  \\
\hline
\end{tabular}_{\quad S^0}
\nonumber \\
+ 
&\begin{tabular}{|c|c|}
\hline
1 & 2   \\
\cline{1-2}
3 &  5  \\
\cline{1-2}
4 &  6  \\
\hline
\end{tabular}_{\quad C}
\otimes
\begin{tabular}{|c|c|c|}
\hline
                1   & 3   & 4    \\
\cline{1-3}  2  &  5 & 6  \\
\hline
\end{tabular}_{\quad S^0}
- 
\begin{tabular}{|c|c|}
\hline
1 & 3   \\
\cline{1-2}
2 &  5  \\
\cline{1-2}
4 &  6  \\
\hline
\end{tabular}_{\quad  C}
\otimes
\begin{tabular}{|c|c|c|}
\hline
                1   & 2   & 4   \\
\cline{1-3} 3  &  5 & 6  \\
\hline
\end{tabular}_{\quad S^0}
\nonumber \\
+ 
&\begin{tabular}{|c|c|}
\hline
1 & 4   \\
\cline{1-2}
2 &  5  \\
\cline{1-2}
3 &  6  \\
\hline
\end{tabular}_{\quad C}
\otimes
\begin{tabular}{|c|c|c|}
\hline
                1   & 2   & 3    \\
\cline{1-3} 4  &  5 & 6  \\
\hline
\end{tabular}_{\quad S^0}
\Big]
\label{eq-CS-3}
\end{align}

\end{itemize}
 
For the isospin part with I=0, the constituent quarks except for two strange quarks of the dibaryon comprise  Young-Yamanouchi basis of Young tableau [2,2] with dimension 2, given by,  
\begin{align}
&\begin{tabular}{|c|c|}
\hline
1 & 2   \\
\cline{1-2}
3 &  4  \\
\cline{1-2}
\end{tabular}
=\frac{1}{\sqrt{12}} I^0_1
\nonumber \\
&=\frac{1}{\sqrt{12}}(2uudd+2dduu-udud-uddu-duud-dudu),
\nonumber \\
&\begin{tabular}{|c|c|}
\hline
1 & 3   \\
\cline{1-2}
2 &  4  \\
\cline{1-2}
\end{tabular}
=\frac{1}{2} I^0_2=\frac{1}{2}(udud-uddu-duud+dudu).
\label{eq-I-0}
\end{align}
Then, we can construct the $\{1234\}\{56\}$ state of the dibaryon with I=0 and S=0 by combining the isospin
state with the color $\otimes$ spin state of Young-Yamanouchi basis [2,2,1,1] ;
\begin{align}
&\{1234\}\{56\}_{[F^{27};I^0, C, S^0]} =\frac{1}{\sqrt{2}}s(5)s(6)\otimes
\nonumber \\
&\Big[\quad\begin{tabular}{|c|c|}
\hline
     1       & 2  \\
\cline{1-2} 
3  &   4 \\
\cline{1-2}
\end{tabular}_{\quad I}
\otimes
\begin{tabular}{|c|c|}
\hline
     1       & 3  \\
\cline{1-2} 
2 &   4  \\
\cline{1-2}
\multicolumn{1}{|c|}{5}  \\
\cline{1-1}
\multicolumn{1}{|c|}{6}  \\
\cline{1-1}
\end{tabular}_{\quad CS}
-\quad\begin{tabular}{|c|c|}
\hline
     1       & 3  \\
\cline{1-2} 
2  &   4 \\
\cline{1-2}
\end{tabular}_{\quad I}
\otimes
\begin{tabular}{|c|c|}
\hline
     1       & 2  \\
\cline{1-2} 
3 &   4  \\
\cline{1-2}
\multicolumn{1}{|c|}{5}  \\
\cline{1-1}
\multicolumn{1}{|c|}{6}  \\
\cline{1-1}
\end{tabular}_{\quad CS}
\Big]
\label{eq-F27-I0}
\end{align}
By $F^{27}$ we mean that this state originally come from a fully antisymmetric state with flavor 27 multiplet,
as the symmetry breaking of flavor SU(3) is imposed on the constituent quarks. The symmetry property is easily perceived in a sense that the linear sum of Young-Yamanouchi [2,2] basis $\otimes$ Young-Yamanouchi [2,2] basis
gives $[1^4]$ basis for particles 1, 2, 3, and 4, and particles 5 and 6 in the color $\otimes$ spin state is antisymmetric due to the positions in the same column.

In addition to the state, $\{1234\}\{56\}_{[F^{27};I^0, C, S^0]}$, there is another state, originally coming from
a fully antisymmetric state with flavor singlet, as the symmetry breaking of flavor SU(3) is imposed on the constituent quarks ;
\begin{align}
&\{1234\}\{56\}_{[F^1;I^0, C, S^0]} =\frac{1}{\sqrt{2}}s(5)s(6)\otimes
\nonumber \\
&\Big[\quad\begin{tabular}{|c|c|}
\hline
     1       & 2  \\
\cline{1-2} 
3  &   4 \\
\cline{1-2}
\end{tabular}_{\quad I}
\otimes
\begin{tabular}{|c|c|c|}
\hline
     1       & 3 & 5  \\
\cline{1-3} 
2 &   4 & 6 \\
\cline{1-3}
\end{tabular}_{\quad CS}
-\quad\begin{tabular}{|c|c|}
\hline
     1       & 3  \\
\cline{1-2} 
2  &   4 \\
\cline{1-2}
\end{tabular}_{\quad I}
\otimes
\begin{tabular}{|c|c|c|}
\hline
     1       & 2 & 5  \\
\cline{1-3} 
3 &   4 & 6 \\
\cline{1-3}
\end{tabular}_{\quad CS}
\Big]
\label{eq-F1-I0}
\end{align}

For the isospin part with I=1, the constituent quarks, apart from the two strange quarks of the dibaryon, comprise the  Young-Yamanouchi basis of Young tableau [3,1] with dimension 3, given by,  
\begin{align}
&\begin{tabular}{|c|c|c|}
\hline
1 & 2  &3 \\
\cline{1-3}
\multicolumn{1}{|c|}{4}  \\
\cline{1-1}
\end{tabular}
=\frac{1}{\sqrt{12}} I^1_1=\frac{1}{\sqrt{12}}(3uuud-uduu-uudu-duuu),
\nonumber \\
&\begin{tabular}{|c|c|c|}
\hline
1 & 2  &4 \\
\cline{1-3}
\multicolumn{1}{|c|}{3}  \\
\cline{1-1}
\end{tabular}
=\frac{1}{\sqrt{6}} I^1_2=\frac{1}{\sqrt{6}}(2uudu-uduu-duuu),
\nonumber \\
&\begin{tabular}{|c|c|c|}
\hline
1 & 3  &4 \\
\cline{1-3}
\multicolumn{1}{|c|}{2}  \\
\cline{1-1}
\end{tabular}
=\frac{1}{\sqrt{2}} I^1_3=\frac{1}{\sqrt{2}}(uduu-duuu).
\label{eq-I-1}
\end{align}
Then, we can construct the $\{1234\}\{56\}$ state of the dibaryon with I=1 and S=0 by combining the isospin
state with the color $\otimes$ spin state of Young-Yamanouchi basis [2,2,1,1] ;
\begin{align}
\{1234\}\{56&\}_{[F^{27};I^1, C, S^0]} =\frac{1}{\sqrt{3}}s(5)s(6)\otimes
\nonumber \\
\Big[
 \quad
&\begin{tabular}{|c|c|c|}
\hline
1 & 2 &3  \\
\cline{1-3}
\multicolumn{1}{|c|}{4}  \\
\cline{1-1}
\end{tabular}_{ I}
\otimes \Big(-\frac{1}{\sqrt{3}} \quad
\begin{tabular}{|c|c|}
\hline
                1   & 4      \\
\cline{1-2} 
 2  &  6  \\
\cline{1-2} 
\multicolumn{1}{|c|}{3}  \\
\cline{1-1} 
\multicolumn{1}{|c|}{5}  \\
\cline{1-1} 
\end{tabular}_{ CS}
+\frac{\sqrt{2}}{\sqrt{3}} \quad
\begin{tabular}{|c|c|}
\hline
                1   & 4      \\
\cline{1-2} 
 2  &  5  \\
\cline{1-2} 
\multicolumn{1}{|c|}{3}  \\
\cline{1-1} 
\multicolumn{1}{|c|}{6}  \\
\cline{1-1} 
\end{tabular}_{ CS}
\Big)
\nonumber \\
- \quad
&\begin{tabular}{|c|c|c|}
\hline
1 & 2 &4  \\
\cline{1-3}
\multicolumn{1}{|c|}{3}  \\
\cline{1-1}
\end{tabular}_{ I}
\otimes \Big(-\frac{1}{\sqrt{3}} \quad
\begin{tabular}{|c|c|}
\hline
                1   & 3      \\
\cline{1-2} 
 2  &  6  \\
\cline{1-2} 
\multicolumn{1}{|c|}{4}  \\
\cline{1-1} 
\multicolumn{1}{|c|}{5}  \\
\cline{1-1} 
\end{tabular}_{ CS}
+\frac{\sqrt{2}}{\sqrt{3}} \quad
\begin{tabular}{|c|c|}
\hline
                1   & 3      \\
\cline{1-2} 
 2  &  5  \\
\cline{1-2} 
\multicolumn{1}{|c|}{4}  \\
\cline{1-1} 
\multicolumn{1}{|c|}{6}  \\
\cline{1-1} 
\end{tabular}_{ CS}
\Big)
\nonumber \\
-  \quad
&\begin{tabular}{|c|c|c|}
\hline
1 & 3 &4  \\
\cline{1-3}
\multicolumn{1}{|c|}{2}  \\
\cline{1-1}
\end{tabular}_{ I}
\otimes \Big(-\frac{1}{\sqrt{3}}  \quad
\begin{tabular}{|c|c|}
\hline
                1   & 2   \\
\cline{1-2} 
 3  &  6  \\
\cline{1-2} 
\multicolumn{1}{|c|}{4}  \\
\cline{1-1} 
\multicolumn{1}{|c|}{5}  \\
\cline{1-1} 
\end{tabular}_{ CS}
+\frac{\sqrt{2}}{\sqrt{3}}  \quad
\begin{tabular}{|c|c|}
\hline
                1   & 2      \\
\cline{1-2} 
 3  &  5  \\
\cline{1-2} 
\multicolumn{1}{|c|}{4}  \\
\cline{1-1} 
\multicolumn{1}{|c|}{6}  \\
\cline{1-1} 
\end{tabular}_{ CS}
\Big)\Big]
\label{eq-F27-I1}
\end{align}
In the same way, the $\{1234\}\{56\}_{[F^{27};I^1, C, S^0]}$ originates from a fully antisymmetric state with flavor 27 multiplet. In this case, the symmetry property for particles 1, 2, 3, and 4 is easily perceived due to the fact that the linear sum of Young-Yamanouchi [3,1] basis $\otimes$ Young-Yamanouchi [2,1,1] basis gives $[1^4]$ basis.
However, the symmetry property of $\{56\}$ between particles 5 and 6 is not easily obtained from the color $\otimes$ spin state of the Young-Yamanouchi [2,2,1,1] basis, since particles 5 and 6 are not positioned in the same column. The symmetry property of the Young-Yamanouchi basis could make it possible to construct $\{56\}$ between particles 5 and 6 ; for example, if we consider the following formula,
\begin{align}
\begin{tabular}{|c|c|}
\hline
                1   & 4      \\
\cline{1-2} 
 2  &  5  \\
\cline{1-2} 
\multicolumn{1}{|c|}{3}  \\
\cline{1-1} 
\multicolumn{1}{|c|}{6}  \\
\cline{1-1} 
\end{tabular}_{ CS}
=\frac{3}{\sqrt{8}}\Big[
(56)
\begin{tabular}{|c|c|}
\hline
                1   & 4      \\
\cline{1-2} 
 2  &  6  \\
\cline{1-2} 
\multicolumn{1}{|c|}{3}  \\
\cline{1-1} 
\multicolumn{1}{|c|}{5}  \\
\cline{1-1} 
\end{tabular}_{ CS}
-\frac{1}{3}
\begin{tabular}{|c|c|}
\hline
                1   & 4      \\
\cline{1-2} 
 2  &  6  \\
\cline{1-2} 
\multicolumn{1}{|c|}{3}  \\
\cline{1-1} 
\multicolumn{1}{|c|}{5}  \\
\cline{1-1} 
\end{tabular}_{ CS}
\Big],
\label{eq-ex-1}
\end{align}
where (56) is a permutation operator between particle 5 and 6, then we see the symmetry property between
5 and 6 by the following formula,
\begin{align}
-\frac{1}{\sqrt{3}}
\begin{tabular}{|c|c|}
\hline
                1   & 4      \\
\cline{1-2} 
 2  &  6  \\
\cline{1-2} 
\multicolumn{1}{|c|}{3}  \\
\cline{1-1} 
\multicolumn{1}{|c|}{5}  \\
\cline{1-1} 
\end{tabular}_{ CS}
+\frac{\sqrt{2}}{\sqrt{3}} 
\begin{tabular}{|c|c|}
\hline
                1   & 4      \\
\cline{1-2} 
 2  &  5  \\
\cline{1-2} 
\multicolumn{1}{|c|}{3}  \\
\cline{1-1} 
\multicolumn{1}{|c|}{6}  \\
\cline{1-1} 
\end{tabular}_{ CS}
=\frac{\sqrt{3}}{2}
\Big[ (56)
\begin{tabular}{|c|c|}
\hline
                1   & 4      \\
\cline{1-2} 
 2  &  6  \\
\cline{1-2} 
\multicolumn{1}{|c|}{3}  \\
\cline{1-1} 
\multicolumn{1}{|c|}{5}  \\
\cline{1-1} 
\end{tabular}_{ CS}
-
\begin{tabular}{|c|c|}
\hline
                1   & 4      \\
\cline{1-2} 
 2  &  6  \\
\cline{1-2} 
\multicolumn{1}{|c|}{3}  \\
\cline{1-1} 
\multicolumn{1}{|c|}{5}  \\
\cline{1-1} 
\end{tabular}_{ CS}
\Big].
\label{eq-ex-2}
\end{align}
Therefore, from the right hand side in Eq.~(\ref{eq-ex-2}), we can show that the left side in Eq.~(\ref{eq-ex-2}) is antisymmetric with respect to the exchange between particle 5 and 6. 

For the isospin part with I=2, the constituent quarks, except for the two strange quarks of the dibaryon, comprise the Young-Yamanouchi bases of Young tableau [4] with dimension 1, given by,  
\begin{align}
&\begin{tabular}{|c|c|c|c|}
\hline
1 & 2  &3 & 4\\
\cline{1-4}
\end{tabular}
= I^2=uuuu.
\label{eq-I-2}
\end{align}

Then, we can construct the $\{1234\}\{56\}$ state of the dibaryon with I=2 and S=0 by combining the isospin
state with the color $\otimes$ spin state of Young-Yamanouchi basis $[1^6]$ ;
\begin{align}
\{1234\}&\{56\}_{[F^{28};I^2, C, S^0]} =
\begin{tabular}{|c|c|c|c|}
\hline
1 & 2 & 3 & 4  \\
\hline
\end{tabular}_{\quad I}
s(5)s(6)\otimes\frac{1}{\sqrt{5}}
\nonumber \\
\Big[
-\quad &\begin{tabular}{|c|c|}
\hline
1 & 2   \\
\cline{1-2}
3 &  4  \\
\cline{1-2}
5 &  6  \\
\hline
\end{tabular}_{\quad C}
\otimes
\begin{tabular}{|c|c|c|}
\hline
                1   & 3   & 5    \\
\cline{1-3} 2  &  4 & 6  \\
\hline
\end{tabular}_{\quad S^0}
+\quad
\begin{tabular}{|c|c|}
\hline
1 & 3  \\
\cline{1-2}
2 &  4  \\
\cline{1-2}
5 &  6  \\
\hline
\end{tabular}_{\quad C}
\otimes
\begin{tabular}{|c|c|c|}
\hline
                1   & 2   & 5    \\
\cline{1-3}  3 &  4 & 6  \\
\hline
\end{tabular}_{\quad S^0}
\nonumber \\
+\quad&\begin{tabular}{|c|c|}
\hline
1 & 2   \\
\cline{1-2}
3 &  5  \\
\cline{1-2}
4 &  6  \\
\hline
\end{tabular}_{\quad C}
\otimes
\begin{tabular}{|c|c|c|}
\hline
                1   & 3   & 4    \\
\cline{1-3}  2  &  5 & 6  \\
\hline
\end{tabular}_{\quad S^0}
 -\quad\begin{tabular}{|c|c|}
\hline
1 & 3   \\
\cline{1-2}
2 &  5  \\
\cline{1-2}
4 &  6  \\
\hline
\end{tabular}_{\quad  C}
\otimes
\begin{tabular}{|c|c|c|}
\hline
                1   & 2   & 4   \\
\cline{1-3} 3  &  5 & 6  \\
\hline
\end{tabular}_{\quad S^0}
\nonumber \\
+\quad&\begin{tabular}{|c|c|}
\hline
1 & 4   \\
\cline{1-2}
2 &  5  \\
\cline{1-2}
3 &  6  \\
\hline
\end{tabular}_{\quad C}
\otimes
\begin{tabular}{|c|c|c|}
\hline
                1   & 2   & 3    \\
\cline{1-3} 4  &  5 & 6  \\
\hline
\end{tabular}_{\quad S^0}
\Big].
\label{eq-F28-I2}
\end{align}
In the same way, the $\{1234\}\{56\}_{[F^{28};I^2, C, S^0]}$ originates from a fully antisymmetric state with flavor 28 multiplet. In this case, the symmetry property is easily understood due to the fact that
the color $\otimes$ spin state of Young-Yamanouchi basis $[1^6]$ is fully antisymmetric among particles 1, 2, 3, 4, 5,
and 6.
 
In addition to the state, $\{1234\}\{56\}_{[F^{28};I^2, C, S^0]}$, there is another state, originally coming from
a fully antisymmetric state with flavor 27 multiplet ;
\begin{align}
\{1234\}&\{56\}_{[F^{27};I^2, C, S^0]} =
\begin{tabular}{|c|c|c|c|}
\hline
1 & 2 & 3 & 4  \\
\hline
\end{tabular}_{\quad I}
s(5)s(6)\otimes
\begin{tabular}{|c|c|}
\hline
                1   & 5      \\
\cline{1-2} 
 2  &  6 \\
\cline{1-2} 
\multicolumn{1}{|c|}{3}  \\
\cline{1-1} 
\multicolumn{1}{|c|}{4}  \\
\cline{1-1} 
\end{tabular}_{ CS}
\label{eq-F27-I2}
\end{align}
In this case, the symmetry property is easily understood due to the fact that the color $\otimes$ spin state of Young-Yamanouchi basis [2,2,1,1] is antisymmetric among particles 1, 2, 3, and 4,
and antisymmetric in the exchange of 5 and 6.

\subsection{Completely antisymmetric flavor $\otimes$ color $\otimes$ spin state of the dibaryon with S=0}

Until now, we investigated the classification and the flavor $\otimes$ color $\otimes$ spin state of the dibaryon, which contains two strange quarks with S=0. From now on, as mentioned in subsection B, we will show that
the isospin $\otimes$ color $\otimes$ spin state of the dibaryon with S=0 found in subsection B can be explicitly
extracted from the completely antisymmetric flavor $\otimes$ color $\otimes$ spin state, in the symmetry breaking of
SU(3) condition. As described in Ref.~\cite{SilvestreBrac:1992yg}, from a point of view of SU(3), the Young tableau of flavor state containing two strange quarks for I=0, I=1, and I=2, is represented as follows :
\begin{itemize}
\item F=1 : 5 basis functions with Young tableau [2,2,2] 
\begin{align}
&\begin{tabular}{c}
$\vert F_1^1 \rangle$=
\end{tabular}
\begin{tabular}{|c|c|}
\hline
1 & 2   \\
\cline{1-2}
3 &  4  \\
\cline{1-2}
5 &  6  \\
\hline
\end{tabular}
\quad
\begin{tabular}{c}
$\vert F_2^1 \rangle$=
\end{tabular}
\begin{tabular}{|c|c|}
\hline
1 & 3   \\
\cline{1-2}
2 &  4  \\
\cline{1-2}
5 &  6  \\
\hline
\end{tabular} 
\quad
\begin{tabular}{c}
$\vert F_3^1 \rangle$=
\end{tabular}
\begin{tabular}{|c|c|}
\hline
1 & 2   \\
\cline{1-2}
3 &  5  \\
\cline{1-2}
4 &  6  \\
\hline
\end{tabular} 
\quad
\begin{tabular}{c}
$\vert F_4^1 \rangle$=
\end{tabular}
\begin{tabular}{|c|c|}
\hline
1 & 3   \\
\cline{1-2}
2 &  5  \\
\cline{1-2}
4 &  6  \\
\hline
\end{tabular}
\nonumber 
\\
&\begin{tabular}{c}
$\vert F_5^1 \rangle$=
\end{tabular}
\begin{tabular}{|c|c|}
\hline
1 & 4   \\
\cline{1-2}
2 &  5  \\
\cline{1-2}
3 &  6  \\
\hline
\end{tabular} 
\quad\quad\quad\quad\quad\quad\quad\quad\quad\quad\quad\quad\quad\quad
\quad\quad\quad\quad
\nonumber 
\end{align}
\item F=27 : 9 basis functions with Young tableau [4,2]
\begin{align}
&\begin{tabular}{c}
$\vert F_1^{27} \rangle$=
\end{tabular}
\begin{tabular}{|c|c|c|c|}
\hline
                  1 & 2 & 3 & 4   \\
\cline{1-4}
\multicolumn{1}{|c|}{5} & \multicolumn{1}{c|}{6}  \\
\cline{1-2}
\end{tabular}
\quad
\begin{tabular}{c}
$\vert F_2^{27} \rangle$=
\end{tabular}
\begin{tabular}{|c|c|c|c|}
\hline
                  1 & 2 & 3 & 5   \\
\cline{1-4}
\multicolumn{1}{|c|}{4} & \multicolumn{1}{c|}{6}  \\
\cline{1-2}
\end{tabular}
\quad
\begin{tabular}{c}
$\vert F_3^{27} \rangle$=
\end{tabular}
\begin{tabular}{|c|c|c|c|}
\hline
                  1 & 2 & 4 & 5   \\
\cline{1-4}
\multicolumn{1}{|c|}{3} & \multicolumn{1}{c|}{6}  \\
\cline{1-2}
\end{tabular}
\nonumber
\\
&\begin{tabular}{c}
$\vert F_4^{27} \rangle$=
\end{tabular}
\begin{tabular}{|c|c|c|c|}
\hline
                  1 & 3 & 4 & 5   \\
\cline{1-4}
\multicolumn{1}{|c|}{2} & \multicolumn{1}{c|}{6}  \\
\cline{1-2}
\end{tabular}
\quad
\begin{tabular}{c}
$\vert F_5^{27} \rangle$=
\end{tabular}
\begin{tabular}{|c|c|c|c|}
\hline
                  1 & 2 & 3 & 6   \\
\cline{1-4}
\multicolumn{1}{|c|}{4} & \multicolumn{1}{c|}{5}  \\
\cline{1-2}
\end{tabular}
\quad
\begin{tabular}{c}
$\vert F_6^{27} \rangle$=
\end{tabular}
\begin{tabular}{|c|c|c|c|}
\hline
                  1 & 2 & 4 & 6   \\
\cline{1-4}
\multicolumn{1}{|c|}{3} & \multicolumn{1}{c|}{5}  \\
\cline{1-2}
\end{tabular}
\nonumber
\\
&\begin{tabular}{c}
$\vert F_7^{27} \rangle$=
\end{tabular}
\begin{tabular}{|c|c|c|c|}
\hline
                  1 & 3 & 4 & 6   \\
\cline{1-4}
\multicolumn{1}{|c|}{2} & \multicolumn{1}{c|}{5}  \\
\cline{1-2}
\end{tabular}
\quad
\begin{tabular}{c}
$\vert F_8^{27} \rangle$=
\end{tabular}
\begin{tabular}{|c|c|c|c|}
\hline
                  1 & 2 & 5 & 6   \\
\cline{1-4}
\multicolumn{1}{|c|}{3} & \multicolumn{1}{c|}{4}  \\
\cline{1-2}
\end{tabular}
\quad
\begin{tabular}{c}
$\vert F_9^{27} \rangle$=
\end{tabular}
\begin{tabular}{|c|c|c|c|}
\hline
                  1 & 3 & 5 & 6   \\
\cline{1-4}
\multicolumn{1}{|c|}{2} & \multicolumn{1}{c|}{4}  \\
\cline{1-2}
\end{tabular}
 \nonumber
\end{align}
\item F=28 : 1 basis function with Young tableau [6] 
\begin{align}
\begin{tabular}{c}
$\vert F^{28} \rangle$=
\end{tabular}
\begin{tabular}{|c|c|c|c|c|c|}
\hline
                  1 & 2 & 3 & 4 & 5 & 6   \\
\hline
\end{tabular}
\nonumber
\end{align}
\end{itemize}
In Appendix A, each of  the flavor state, which will be used to understand our procedure, will be presented in detail.

In the same way as described in the previous paper~\cite{Park:2015nha}, we can construct the fully antisymmetric flavor $\otimes$ color $\otimes$ spin state of the dibaryon with S=0, by combining the flavor state with the color $\otimes$ spin state, which is obtained from CS
couping scheme in subsection B. For each of flavor state, the fully antisymmetric flavor $\otimes$ color $\otimes$ spin state is given as follows :
\begin{align}
&\vert [F^{27},C,S^0] \rangle
=\frac{1}{\sqrt{9}}
\Big[
\nonumber \\
&\begin{tabular}{|c|c|c|c|}
\hline
                  1 & 2 & 3 & 4   \\
\cline{1-4}
\multicolumn{1}{|c|}{5} & \multicolumn{1}{c|}{6}  \\
\cline{1-2}
\end{tabular}_F
\otimes
\begin{tabular}{|c|c|}
\hline
1 & 5   \\
\cline{1-2}
2 &  6  \\
\cline{1-2}
\multicolumn{1}{|c|}{3}  \\
\cline{1-1}
\multicolumn{1}{|c|}{4}  \\
\cline{1-1}
\end{tabular}_{CS} 
-\quad
\begin{tabular}{|c|c|c|c|}
\hline
                  1 & 2 & 3 & 5   \\
\cline{1-4}
\multicolumn{1}{|c|}{4} & \multicolumn{1}{c|}{6}  \\
\cline{1-2}
\end{tabular}_F
\otimes
\begin{tabular}{|c|c|}
\hline
1 & 4   \\
\cline{1-2}
2 &  6  \\
\cline{1-2}
\multicolumn{1}{|c|}{3}  \\
\cline{1-1}
\multicolumn{1}{|c|}{5}  \\
\cline{1-1}
\end{tabular}_{CS}
+
\nonumber \\
&\begin{tabular}{|c|c|c|c|}
\hline
                  1 & 2 & 4 & 5   \\
\cline{1-4}
\multicolumn{1}{|c|}{3} & \multicolumn{1}{c|}{6}  \\
\cline{1-2}
\end{tabular}_F
\otimes
\begin{tabular}{|c|c|}
\hline
1 & 3   \\
\cline{1-2}
2 &  6  \\
\cline{1-2}
\multicolumn{1}{|c|}{4}  \\
\cline{1-1}
\multicolumn{1}{|c|}{5}  \\
\cline{1-1}
\end{tabular}_{CS}
-\quad
\begin{tabular}{|c|c|c|c|}
\hline
                  1 & 3 & 4 & 5   \\
\cline{1-4}
\multicolumn{1}{|c|}{2} & \multicolumn{1}{c|}{6}  \\
\cline{1-2}
\end{tabular}_F
\otimes
\begin{tabular}{|c|c|}
\hline
1 & 2   \\
\cline{1-2}
3 &  6  \\
\cline{1-2}
\multicolumn{1}{|c|}{4}  \\
\cline{1-1}
\multicolumn{1}{|c|}{5}  \\
\cline{1-1}
\end{tabular}_{CS}
 +
\nonumber \\ 
&\begin{tabular}{|c|c|c|c|}
\hline
                  1 & 2 & 3 & 6   \\
\cline{1-4}
\multicolumn{1}{|c|}{4} & \multicolumn{1}{c|}{5}  \\
\cline{1-2}
\end{tabular}_F
\otimes
\begin{tabular}{|c|c|}
\hline
1 & 4   \\
\cline{1-2}
2 &  5  \\
\cline{1-2}
\multicolumn{1}{|c|}{3}  \\
\cline{1-1}
\multicolumn{1}{|c|}{6}  \\
\cline{1-1}
\end{tabular}_{CS} 
-\quad
\begin{tabular}{|c|c|c|c|}
\hline
                  1 & 2 & 4 & 6   \\
\cline{1-4}
\multicolumn{1}{|c|}{3} & \multicolumn{1}{c|}{5}  \\
\cline{1-2}
\end{tabular}_F
\otimes
\begin{tabular}{|c|c|}
\hline
1 & 3   \\
\cline{1-2}
2 &  5  \\
\cline{1-2}
\multicolumn{1}{|c|}{4}  \\
\cline{1-1}
\multicolumn{1}{|c|}{6}  \\
\cline{1-1}
\end{tabular}_{CS}
+
\nonumber \\ 
&\begin{tabular}{|c|c|c|c|}
\hline
                  1 & 3 & 4 & 6   \\
\cline{1-4}
\multicolumn{1}{|c|}{2} & \multicolumn{1}{c|}{5}  \\
\cline{1-2}
\end{tabular}_F
\otimes
\begin{tabular}{|c|c|}
\hline
1 & 2   \\
\cline{1-2}
3 &  5  \\
\cline{1-2}
\multicolumn{1}{|c|}{4}  \\
\cline{1-1}
\multicolumn{1}{|c|}{6}  \\
\cline{1-1}
\end{tabular}_{CS}
+\quad
\begin{tabular}{|c|c|c|c|}
\hline
                  1 & 2 & 5 & 6   \\
\cline{1-4}
\multicolumn{1}{|c|}{3} & \multicolumn{1}{c|}{4}  \\
\cline{1-2}
\end{tabular}_F
\otimes
\begin{tabular}{|c|c|}
\hline
1 & 3  \\
\cline{1-2}
2 &  4  \\
\cline{1-2}
\multicolumn{1}{|c|}{5}  \\
\cline{1-1}
\multicolumn{1}{|c|}{6}  \\
\cline{1-1}
\end{tabular}_{CS}
-
\nonumber \\ 
&\begin{tabular}{|c|c|c|c|}
\hline
                  1 & 3 & 5 & 6   \\
\cline{1-4}
\multicolumn{1}{|c|}{2} & \multicolumn{1}{c|}{4}  \\
\cline{1-2}
\end{tabular}_F
\otimes
\begin{tabular}{|c|c|}
\hline
1 & 2   \\
\cline{1-2}
3 &  4  \\
\cline{1-2}
\multicolumn{1}{|c|}{5}  \\
\cline{1-1}
\multicolumn{1}{|c|}{6}  \\
\cline{1-1}
\end{tabular}_{CS}
\Big].
\label{eq-F27-C1}
\end{align} 
Eq.~(\ref{eq-F27-C1}) is a fully antisymmetric state for I=0, I=1 and I=2, in which  the flavor 27 multiplet lies.
\begin{align}
&\vert [F^{1},C,S^0] \rangle
=\frac{1}{\sqrt{5}}
\Big[
\nonumber \\
&\begin{tabular}{|c|c|}
\hline
1 & 2   \\
\cline{1-2}
3 &  4  \\
\cline{1-2}
5 &  6  \\
\hline
\end{tabular}_{\quad F}
\otimes
\begin{tabular}{|c|c|c|}
\hline
                1   & 3   & 5   \\
\cline{1-3}  2  &  4 & 6  \\
\hline
\end{tabular}_{\quad CS}
-\quad
\begin{tabular}{|c|c|}
\hline
1 & 3  \\
\cline{1-2}
2 &  4  \\
\cline{1-2}
5 &  6  \\
\hline
\end{tabular}_{\quad F}
\otimes
\begin{tabular}{|c|c|c|}
\hline
                1   & 2   & 5    \\
\cline{1-3} 3 &  4 & 6  \\
\hline
\end{tabular}_{\quad CS}
-
\nonumber \\
&\begin{tabular}{|c|c|}
\hline
1 & 2   \\
\cline{1-2}
3 &  5  \\
\cline{1-2}
4 &  6  \\
\hline
\end{tabular}_{\quad F}
\otimes
\begin{tabular}{|c|c|c|}
\hline
                1   & 3   & 4    \\
\cline{1-3}  2  &  5 & 6  \\
\hline
\end{tabular}_{\quad CS}
-\quad
\begin{tabular}{|c|c|}
\hline
1 & 3   \\
\cline{1-2}
2 &  5  \\
\cline{1-2}
4 &  6  \\
\hline
\end{tabular}_{\quad  F}
\otimes
\begin{tabular}{|c|c|c|}
\hline
                1   & 2   & 4    \\
\cline{1-3}  3  &  5 & 6  \\
\hline
\end{tabular}_{\quad CS}
-
\nonumber \\
&\begin{tabular}{|c|c|}
\hline
1 & 4   \\
\cline{1-2}
2 &  5  \\
\cline{1-2}
3 &  6  \\
\hline
\end{tabular}_{\quad F}
\otimes
\begin{tabular}{|c|c|c|}
\hline
                1   & 2   & 3    \\
\cline{1-3} 4  &  5 & 6  \\
\hline
\end{tabular}_{\quad CS}
\Big]. 
\label{eq-F1-C2}
\end{align}
Eq.~(\ref{eq-F1-C2}) is a fully antisymmetric state for I=0, in which the flavor singlet state lies.
\begin{align}
&\vert [F^{28},C,S^0] \rangle
=
\begin{tabular}{|c|c|c|c|c|c|}
\hline
1 & 2 & 3 & 4 & 5 & 6 \\
\hline
\end{tabular}_{\quad F}
\otimes
\frac{1}{\sqrt{5}}
\Big[
\nonumber \\
&-\quad \begin{tabular}{|c|c|}
\hline
1 & 2   \\
\cline{1-2}
3 &  4  \\
\cline{1-2}
5 &  6  \\
\hline
\end{tabular}_{\quad C}
\otimes
\begin{tabular}{|c|c|c|}
\hline
                1   & 3   & 5    \\
\cline{1-3} 2  &  4 & 6  \\
\hline
\end{tabular}_{\quad S^0}
+\quad
\begin{tabular}{|c|c|}
\hline
1 & 3  \\
\cline{1-2}
2 &  4  \\
\cline{1-2}
5 &  6  \\
\hline
\end{tabular}_{\quad C}
\otimes
\begin{tabular}{|c|c|c|}
\hline
                1   & 2   & 5    \\
\cline{1-3}  3 &  4 & 6  \\
\hline
\end{tabular}_{\quad S^0}
\nonumber \\
&+\quad
\begin{tabular}{|c|c|}
\hline
1 & 2   \\
\cline{1-2}
3 &  5  \\
\cline{1-2}
4 &  6  \\
\hline
\end{tabular}_{\quad C}
\otimes
\begin{tabular}{|c|c|c|}
\hline
                1   & 3   & 4    \\
\cline{1-3}  2  &  5 & 6  \\
\hline
\end{tabular}_{\quad S^0}
-\quad
\begin{tabular}{|c|c|}
\hline
1 & 3   \\
\cline{1-2}
2 &  5  \\
\cline{1-2}
4 &  6  \\
\hline
\end{tabular}_{\quad  C}
\otimes
\begin{tabular}{|c|c|c|}
\hline
                1   & 2   & 4   \\
\cline{1-3} 3  &  5 & 6  \\
\hline
\end{tabular}_{\quad S^0}
\nonumber \\
&+\quad
\begin{tabular}{|c|c|}
\hline
1 & 4   \\
\cline{1-2}
2 &  5  \\
\cline{1-2}
3 &  6  \\
\hline
\end{tabular}_{\quad C}
\otimes
\begin{tabular}{|c|c|c|}
\hline
                1   & 2   & 3    \\
\cline{1-3} 4  &  5 & 6  \\
\hline
\end{tabular}_{\quad S^0}
\Big].  
\label{eq-F28-C3}
\end{align}
Eq.~(\ref{eq-F28-C3}) is a fully antisymmetric state for I=2, in which the flavor 28 multiplet state lies.

In order to consider the symmetry breaking of SU(3) in these fully antisymmetric states with respect to isospin I,
we need to restrict the position of the two strange quarks to any two  among the possible 15 (${}_6C_2$) places. In our work,
the fixing of the positions of the two strange quarks onto the fifth and sixth is performed by taking the cases where the strange quark resides in other positions to zero; s(1), s(2), s(3), and s(4) $\rightarrow$ 0. 
Then, Eq.~(\ref{eq-F27-C1}) exactly becomes equal to  1/$\sqrt{15}$ $\{1234\}\{56\}_{[F^{27};I^0, C, S^0]}$ for I=0 in Eq.~(\ref{eq-F27-I0}) under the condition, because the other terms of flavor 27 multiplet except for
 $\vert F^{27}_8 \rangle$ and $\vert F^{27}_9 \rangle$ vanish as shown in Appendix A. The  factor 1/$\sqrt{15}$, which will become clear later, is related with the symmetry property of the fully antisymmetric state.  As another example for I=1,  Eq.~(\ref{eq-F27-C1}) exactly becomes equal to the 1/$\sqrt{15}$ $\{1234\}\{56\}_{[F^{27};I^1, C, S^0]}$ in Eq.~(\ref{eq-F27-I1}), because $\vert F^{27}_1 \rangle$, $\vert F^{27}_8 \rangle$, and $\vert F^{27}_9 \rangle$ terms of flavor 27 multiplet vanish as shown in Appendix A. With this approach, we find that the completely antisymmetric flavor $\otimes$ color $\otimes$ spin state in Eq.~(\ref{eq-F27-C1}), Eq.~(\ref{eq-F1-C2}), and Eq.~(\ref{eq-F28-C3}) becomes the $\{1234\}\{56\}$ states obtained in subsection B with respect to I. 

We need to investigate more to make the completely antisymmetric flavor $\otimes$ color  $\otimes$ spin state in terms of $\{1234\}\{56\}$ state, prior to finishing  this subsection. According to the way of fixing two strange quarks, there are 14 more partially  antisymmetric states in addition to $\{1234\}\{56\}$ state, where the two strange quarks is located on the fifth and sixth position in the $\{1234\}\{56\}$ states. These states are given as follows; $\{3456\}\{12\}$, $\{2456\}\{13\}$, $\{2356\}\{14\}$, $\{2346\}\{15\}$, $\{2345\}\{16\}$, $\{1456\}\{23\}$, $\{1356\}\{24\}$, $\{1346\}\{25\}$, $\{1345\}\{26\}$, $\{1256\}\{34\}$, $\{1246\}\{35\}$, $\{1245\}\{36\}$, $\{1236\}\{45\}$, $\{1235\}\{46\}$, in addition to $\{1234\}\{56\}$. \\
These states are orthonormal to each other, because one has at least a strange quark in a different position, in contrast to
the others. Moreover, the states for I is also obtained from the completely antisymmetric flavor $\otimes$ color  $\otimes$ spin state, by performing the same procedure with two strange quarks in different positions presented above.  For any $[1^6]$ with respect to I, the completely antisymmetric state can be explicitly expressed by, 

\begin{align}
[1^6]=\frac{1}{\sqrt{15}}
\Big[ &\{3456\}\{12\} +\{2456\}\{13\} +\{2356\}\{14\} +
\nonumber \\
&\{2346\}\{15\} +\{2345\}\{16\} +\{1456\}\{23\} +
\nonumber \\
&\{1356\}\{24\} +\{1346\}\{25\} +\{1345\}\{26\} +
\nonumber \\
&\{1256\}\{34\} +\{1246\}\{35\} +\{1245\}\{36\} +
\nonumber \\
&\{1236\}\{45\} +\{1235\}\{46\} +\{1234\}\{56\}\Big].
\label{eq-antisymmetry}
\end{align}

As interesting points, we find that the other states  in Eq.~(\ref{eq-antisymmetry}) except for $\{1234\}\{56\}$ state  can be directly obtained from the $\{1234\}\{56\}$ state, by using Eq.~(\ref{eq-permutation}), which are identities obtained from Eq.~(\ref{eq-antisymmetry}) itself.  
In  Eq.~(\ref{eq-permutation}), ($ij$) is the permutation operator that exchange between $i$ and $j$. From Eq.~(\ref{eq-permutation}), we can introduce a formula so as to simplify our arguments:   
we define the following formula from the fact that the $\{1234\}\{56\}$ state can be extracted from the completely antisymmetric $[1^6]$ state by taking  s(1), s(2), s(3), and s(4) $\rightarrow$ 0 so that 

\begin{align}
&\lim_{s(1), s(2), s(3), s(4) \to 0} [1^6] \equiv \frac{1}{\sqrt{15}}\{1234\}\{56\} , \nonumber \\
&\lim_{s(1), s(2), s(3), s(5) \to 0} [1^6] \equiv \frac{1}{\sqrt{15}}\{1235\}\{46\} , \nonumber \\
&\lim_{s(1), s(2), s(3), s(6) \to 0} [1^6] \equiv \frac{1}{\sqrt{15}}\{1236\}\{45\} , \nonumber \\
&....., \nonumber \\
&\lim_{s(3), s(4), s(5), s(6) \to 0} [1^6] \equiv \frac{1}{\sqrt{15}}\{3456\}\{12\}.
\label{eq-definition-1}
\end{align}
The acting of the permutation operator ($ij$) of the above equation  on both of sides can be defined so as to satisfy the Eq.~(\ref{eq-permutation}),
given by,

\begin{align}
&(ij)\lim_{s(1), s(2), s(3), s(4) \to 0} [1^6] \equiv \nonumber \\
&\lim_{s((ij)(1)), s((ij)(2)), s((ij)(3)), s((ij)(4)) \to 0}(ij)[1^6] \equiv \nonumber \\
&\lim_{s((ij)(1)), s((ij)(2)), s((ij)(3)), s((ij)(4)) \to 0}(-)[1^6] = \nonumber \\
&\frac{1}{\sqrt{15}}(ij)\{1234\}\{56\}, 
\label{eq-definition-2}
\end{align}
where the minus sign appearing in the third line of the Eq.~(\ref{eq-definition-2}) is due to the fully antisymmetry property of $[1^6]$ 
in acting any permutation operator.
When we apply (15)(26) on both sides of the first line in Eq.~(\ref{eq-definition-1}), we have the same equation to satisfy the  Eq.~(\ref{eq-permutation}) ;

\begin{align}
&(15)(26)\lim_{s(1), s(2), s(3), s(4) \to 0} [1^6] = \nonumber \\
&\lim_{s(5), s(6), s(3), s(4) \to 0}(15)(26)[1^6] = \nonumber \\
&\lim_{s(5), s(6), s(3), s(4) \to 0}[1^6] =\frac{1}{\sqrt{15}}\{3456\}\{12\}.  
\label{eq-general}
\end{align}
Therefore, we find that $\{3456\}\{12\}$ = (15)(26) $\{1234\}\{56\}$. 

\begin{widetext}\allowdisplaybreaks
\begin{align} 
&\{3456\}\{12\} = (15)(26) \{1234\}\{56\}, \quad
\{2456\}\{13\} = (15)(36) \{1234\}\{56\}, \quad
\{2356\}\{14\} = (15)(46) \{1234\}\{56\}, \nonumber \\
&\{2346\}\{15\} =-(16) \{1234\}\{56\}, \quad \quad
\{2345\}\{16\} =-(15) \{1234\}\{56\}, \quad \quad
\{1456\}\{23\} =(25)(36) \{1234\}\{56\}, \nonumber \\
&\{1356\}\{24\} =(25)(46) \{1234\}\{56\},  \quad
\{1346\}\{25\} =-(26)\{1234\}\{56\}, \quad \quad
\{1345\}\{26\} =-(25)\{1234\}\{56\},  \nonumber \\
&\{1256\}\{34\} =(35)(46)\{1234\}\{56\}, \quad
\{1246\}\{35\} =-(36)\{1234\}\{56\}, \quad \quad
\{1245\}\{36\} =-(35)\{1234\}\{56\}, \nonumber \\
&\{1236\}\{45\} =-(46)\{1234\}\{56\},  \quad \quad
\{1235\}\{46\} =-(45)\{1234\}\{56\}.
\label{eq-permutation}
\end{align}
\end{widetext}

Using Eq.~(\ref{eq-definition-2}), it is easily found that the right hand side of the Eq.~(\ref{eq-antisymmetry})
is fully antisymmetric. Also, it should be noted that the expectation value of $-{\sum}_{i<j}^{N=6}\lambda_i^c\lambda_j^c{\sigma}_i\cdot{\sigma}_j$ in terms of $[1^6]$ state with respect to I is simply  reduced to that in terms of $\{1234\}\{56\}$ state. For example, we have the following equation for I=0, coming from flavor singlet;

\begin{align}
&\langle-{\sum}_{i<j}^{N=6}\lambda_i^c\lambda_j^c{\sigma}_i\cdot{\sigma}_j \rangle_{\vert [F^{1},C,S^0] \rangle}=
\frac{1}{15}\Big[
\nonumber \\
&\langle-{\sum}_{i<j}^{N=6}\lambda_i^c\lambda_j^c{\sigma}_i\cdot{\sigma}_j \rangle_{\{3456\}\{12\}_{[F^{1};I^0, C, S^0]}}+
\nonumber \\
&\langle-{\sum}_{i<j}^{N=6}\lambda_i^c\lambda_j^c{\sigma}_i\cdot{\sigma}_j \rangle_{\{2456\}\{13\}_{[F^{1};I^0, C, S^0]}}+
\nonumber \\
&....,+ \nonumber \\
&\langle-{\sum}_{i<j}^{N=6}\lambda_i^c\lambda_j^c{\sigma}_i\cdot{\sigma}_j \rangle_{\{1234\}\{56\}_{[F^{1};I^0, C, S^0]}}
\Big]=\nonumber \\
&\langle-{\sum}_{i<j}^{N=6}\lambda_i^c\lambda_j^c{\sigma}_i\cdot{\sigma}_j \rangle_{\{1234\}\{56\}_{[F^{1};I^0, C, S^0]}}
\label{eq-expectation}
\end{align}

\section{ Calculation of the expectation value of hyperfine potential}

In this section, we calculate the expectation value of hyperfine potential of the dibaryon in terms of isospin $\otimes$ color  $\otimes$ spin state found in section III. Even though the expectation value of hyperfine potential of the dibaryon can be directly obtained from the full flavor $\otimes$ color $\otimes$ spin wave function of the state, the symmetry property of the state enable us to approach  the expectation value of hyperfine potential of the dibaryon in terms of fully antisymmetric state mentioned in subsection C of section III, namely, in terms of the isospin $\otimes$ color  $\otimes$ spin state. In calculating the expectation value of hyperfine potential of the dibaryon, it is much convenient to use the well-established formula~\cite{Aerts:1977rw}, which is applicable for  fully antisymmetric state with SU(2) flavor, given by,
 
\begin{align}
-&{\sum}_{i<j}^{N}\lambda_i^c\lambda_j^c{\sigma}_i\cdot{\sigma}_j = \nonumber\\                                                                                                                     
[&\frac{4}{3}N(N-6)+4I(I+1)+\frac{4}{3}S(S+1)+2C_c],
\label{eq-magic}
\end{align}
where $N$ is the total number of quarks in a system, and $C_c$=$\frac{1}{4}\lambda^c\lambda^c$, that is,
the first kind of Casimir operator of color SU(3) in the system of N quarks. To apply this formula to the $\{1234\}\{56\}$ state
found in subsection C of section III, we need to know the color state among particles 1, 2, 3, and 4, and between 5 and 6,
because both of the color states are not in  color singlet states. We can re-express the $\{1234\}\{56\}$ state for I, by considering
another way of constructing the $\{1234\}\{56\}$ state;   
\begin{widetext}\allowdisplaybreaks
\begin{align}
\vert [C, I^0, S^0]_1 \rangle=\frac{1}{\sqrt{3}}\Big[
&\begin{tabular}{|c|c|}
\hline
     1       & 2  \\
\cline{1-2} 
\multicolumn{1}{|c|}{3} &\multicolumn{1}{c}{5}  \\
\cline{1-1}
\multicolumn{1}{|c|}{4} &\multicolumn{1}{c}{6}  \\
\cline{1-1}
\end{tabular}_C
\otimes(-\frac{1}{2}
\begin{tabular}{|c|c|}
\hline
     1       & 2  \\
\cline{1-2} 
3 &      4  \\
\cline{1-2}
\end{tabular}_{\quad I}
\otimes
\begin{tabular}{|c|c|c|}
\hline
     1       & 3  &4   \\
\cline{1-3} 
\multicolumn{1}{|c|}{2} &\multicolumn{1}{c}{5} &\multicolumn{1}{c}{6} \\
\cline{1-1}
\end{tabular}_{S^0}
+\frac{1}{\sqrt{2}}
\begin{tabular}{|c|c|}
\hline
     1       & 3  \\
\cline{1-2} 
2 &      4  \\
\cline{1-2}
\end{tabular}_{\quad I}
\otimes
\begin{tabular}{|c|c|c|}
\hline
     1       & 2  & 3 \\
\cline{1-3} 
\multicolumn{1}{|c|}{4} &\multicolumn{1}{c}{5} &\multicolumn{1}{c}{6} \\
\cline{1-1}
\end{tabular}_{S^0}
-\frac{1}{2}
\begin{tabular}{|c|c|}
\hline
     1       & 3  \\
\cline{1-2} 
2 &      4  \\
\cline{1-2}
\end{tabular}_{\quad I}
\otimes
\begin{tabular}{|c|c|c|}
\hline
     1       & 2  & 4 \\
\cline{1-3} 
\multicolumn{1}{|c|}{3} &\multicolumn{1}{c}{5} &\multicolumn{1}{c}{6} \\
\cline{1-1}
\end{tabular}_{S^0}
)  \nonumber \\
-&\begin{tabular}{|c|c|}
\hline
     1       & 3  \\
\cline{1-2} 
\multicolumn{1}{|c|}{2} &\multicolumn{1}{c}{5}  \\
\cline{1-1}
\multicolumn{1}{|c|}{4} &\multicolumn{1}{c}{6}  \\
\cline{1-1}
\end{tabular}_C
\otimes(\frac{1}{\sqrt{2}}
\begin{tabular}{|c|c|}
\hline
     1       & 2  \\
\cline{1-2} 
3 &      4  \\
\cline{1-2}
\end{tabular}_{\quad I}
\otimes
\begin{tabular}{|c|c|c|}
\hline
     1       & 2  &3   \\
\cline{1-3} 
\multicolumn{1}{|c|}{4} &\multicolumn{1}{c}{5} &\multicolumn{1}{c}{6} \\
\cline{1-1}
\end{tabular}_{S^0}
+\frac{1}{2}
\begin{tabular}{|c|c|}
\hline
     1       & 2  \\
\cline{1-2} 
3 &      4  \\
\cline{1-2}
\end{tabular}_{\quad I}
\otimes
\begin{tabular}{|c|c|c|}
\hline
     1       & 2  & 4 \\
\cline{1-3} 
\multicolumn{1}{|c|}{3} &\multicolumn{1}{c}{5} &\multicolumn{1}{c}{6} \\
\cline{1-1}
\end{tabular}_{S^0}
-\frac{1}{2}
\begin{tabular}{|c|c|}
\hline
     1       & 3  \\
\cline{1-2} 
2 &      4  \\
\cline{1-2}
\end{tabular}_{\quad I}
\otimes
\begin{tabular}{|c|c|c|}
\hline
     1       & 3  & 4 \\
\cline{1-3} 
\multicolumn{1}{|c|}{2} &\multicolumn{1}{c}{5} &\multicolumn{1}{c}{6} \\
\cline{1-1}
\end{tabular}_{S^0}
)  \nonumber \\
+&\begin{tabular}{|c|c|}
\hline
     1       & 4  \\
\cline{1-2} 
\multicolumn{1}{|c|}{2} &\multicolumn{1}{c}{5}  \\
\cline{1-1}
\multicolumn{1}{|c|}{3} &\multicolumn{1}{c}{6}  \\
\cline{1-1}
\end{tabular}_C
\otimes(\frac{1}{\sqrt{2}}
\begin{tabular}{|c|c|}
\hline
     1       & 2  \\
\cline{1-2} 
3 &      4  \\
\cline{1-2}
\end{tabular}_{\quad I}
\otimes
\begin{tabular}{|c|c|c|}
\hline
     1       & 2  & 4 \\
\cline{1-3} 
\multicolumn{1}{|c|}{3} &\multicolumn{1}{c}{5} &\multicolumn{1}{c}{6} \\
\cline{1-1}
\end{tabular}_{S^0}
+\frac{1}{\sqrt{2}}
\begin{tabular}{|c|c|}
\hline
     1       & 3  \\
\cline{1-2} 
2 &      4  \\
\cline{1-2}
\end{tabular}_{\quad I}
\otimes
\begin{tabular}{|c|c|c|}
\hline
     1       & 3  & 4 \\
\cline{1-3} 
\multicolumn{1}{|c|}{2} &\multicolumn{1}{c}{5} &\multicolumn{1}{c}{6} \\
\cline{1-1}
\end{tabular}_{S^0}
) \Big]\otimes s(5)s(6)
\label{eq-I0-1}
\end{align}
\begin{align}
\vert [C, I^0, S^0]_2 \rangle=\frac{1}{\sqrt{2}}\Big[
&\begin{tabular}{|c|c|}
\hline
     1       & 2  \\
\cline{1-2} 
3  &   4 \\
\cline{1-2}
\multicolumn{1}{c}{5} &\multicolumn{1}{c}{6}  \\
\end{tabular}_C
\otimes(-\frac{1}{\sqrt{2}}
\begin{tabular}{|c|c|}
\hline
     1       & 2  \\
\cline{1-2} 
3 &      4  \\
\cline{1-2}
\end{tabular}_{\quad I}
\otimes
\begin{tabular}{|c|c|c|}
\cline{1-2} 
     1       & 3 & \multicolumn{1}{c}{5} \\
\cline{1-2} 
2  &   4   & \multicolumn{1}{c}{6} \\
\cline{1-2}
\end{tabular}_{S^0}
-\frac{1}{\sqrt{2}}
\begin{tabular}{|c|c|}
\hline
     1       & 3  \\
\cline{1-2} 
2 &      4  \\
\cline{1-2}
\end{tabular}_{\quad I}
\otimes
\begin{tabular}{|c|c|c|}
\cline{1-2} 
     1       & 2 & \multicolumn{1}{c}{5} \\
\cline{1-2} 
3  &   4   & \multicolumn{1}{c}{6} \\
\cline{1-2}
\end{tabular}_{S^0}
)\nonumber \\
-&\begin{tabular}{|c|c|}
\hline
     1       & 3  \\
\cline{1-2} 
2  &   4 \\
\cline{1-2}
\multicolumn{1}{c}{5} &\multicolumn{1}{c}{6}  \\
\end{tabular}_C
\otimes(\frac{1}{\sqrt{2}}
\begin{tabular}{|c|c|}
\hline
     1       & 2  \\
\cline{1-2} 
3 &      4  \\
\cline{1-2}
\end{tabular}_{\quad I}
\otimes
\begin{tabular}{|c|c|c|}
\cline{1-2} 
     1       & 2 & \multicolumn{1}{c}{5} \\
\cline{1-2} 
3  &   4   & \multicolumn{1}{c}{6} \\
\cline{1-2}
\end{tabular}_{S^0}
-\frac{1}{\sqrt{2}}
\begin{tabular}{|c|c|}
\hline
     1       & 3  \\
\cline{1-2} 
2 &      4  \\
\cline{1-2}
\end{tabular}_{\quad I}
\otimes
\begin{tabular}{|c|c|c|}
\cline{1-2} 
     1       & 3 & \multicolumn{1}{c}{5} \\
\cline{1-2} 
2  &   4   & \multicolumn{1}{c}{6} \\
\cline{1-2}
\end{tabular}_{S^0}
)\Big]\otimes s(5)s(6)
\label{eq-I0-2}
\end{align}
\begin{align}
&\vert [C, I^1, S^0] \rangle =\frac{1}{\sqrt{3}} \otimes s(5)s(6)
\Big[
\begin{tabular}{|c|c|c|}
\hline
1 & 2 &3  \\
\cline{1-3}
\multicolumn{1}{|c|}{4}  \\
\cline{1-1}
\end{tabular}_{ I}
\otimes \Big(\frac{2}{\sqrt{6}}
\begin{tabular}{|c|c|}
\hline
     1       & 4  \\
\cline{1-2} 
2  & \multicolumn{1}{c}{5}   \\
\cline{1-1}
\multicolumn{1}{|c|}{3} &\multicolumn{1}{c}{6}  \\
\cline{1-1}
\end{tabular}_C
\otimes 
\begin{tabular}{|c|c|c|}
\cline{1-3} 
     1       & 2 &  \multicolumn{1}{|c|}{3} \\
\cline{1-3} 
4 & \multicolumn{1}{c}{5} & \multicolumn{1}{c}{6} \\
\cline{1-1}
\end{tabular}_{S^0}
+ 
\frac{1}{\sqrt{6}}
\begin{tabular}{|c|c|}
\hline
     1       & 3  \\
\cline{1-2} 
2  & \multicolumn{1}{c}{5}    \\
\cline{1-1}
\multicolumn{1}{|c|}{4} &\multicolumn{1}{c}{6}  \\
\cline{1-1}
\end{tabular}_C
\otimes 
\begin{tabular}{|c|c|c|}
\cline{1-3} 
     1       & 2 &  \multicolumn{1}{|c|}{4} \\
\cline{1-3} 
3 & \multicolumn{1}{c}{5} & \multicolumn{1}{c}{6} \\
\cline{1-1}
\end{tabular}_{S^0}
- 
\frac{1}{\sqrt{6}}
\begin{tabular}{|c|c|}
\hline
     1       & 2  \\
\cline{1-2} 
3 & \multicolumn{1}{c}{5}   \\
\cline{1-1}
\multicolumn{1}{|c|}{4} &\multicolumn{1}{c}{6}  \\
\cline{1-1}
\end{tabular}_C
\otimes 
\begin{tabular}{|c|c|c|}
\cline{1-3} 
     1       & 3 &  \multicolumn{1}{|c|}{4} \\
\cline{1-3} 
2 & \multicolumn{1}{c}{5} & \multicolumn{1}{c}{6} \\
\cline{1-1}
\end{tabular}_{S^0}
\Big)
\nonumber \\
- 
&\begin{tabular}{|c|c|c|}
\hline
1 & 2 &4  \\
\cline{1-3}
\multicolumn{1}{|c|}{3}  \\
\cline{1-1}
\end{tabular}_{ I}
\otimes \Big(\frac{1}{\sqrt{6}}
\begin{tabular}{|c|c|}
\hline
     1       & 4  \\
\cline{1-2} 
2  & \multicolumn{1}{c}{5}    \\
\cline{1-1}
\multicolumn{1}{|c|}{3} &\multicolumn{1}{c}{6}  \\
\cline{1-1}
\end{tabular}_C
\otimes 
\begin{tabular}{|c|c|c|}
\cline{1-3} 
     1       & 2 &  \multicolumn{1}{|c|}{4} \\
\cline{1-3} 
3 & \multicolumn{1}{c}{5} & \multicolumn{1}{c}{6} \\
\cline{1-1}
\end{tabular}_{S^0}
- 
\frac{1}{\sqrt{6}}
\begin{tabular}{|c|c|}
\hline
     1       & 3  \\
\cline{1-2} 
2  &  \multicolumn{1}{c}{5}  \\
\cline{1-1}
\multicolumn{1}{|c|}{4} &\multicolumn{1}{c}{6}  \\
\cline{1-1}
\end{tabular}_C
\otimes 
\begin{tabular}{|c|c|c|}
\cline{1-3} 
     1       & 2 &  \multicolumn{1}{|c|}{3} \\
\cline{1-3} 
4 & \multicolumn{1}{c}{5} & \multicolumn{1}{c}{6} \\
\cline{1-1}
\end{tabular}_{S^0}
+ 
\frac{1}{\sqrt{3}}
\begin{tabular}{|c|c|}
\hline
     1       & 3  \\
\cline{1-2} 
2  & \multicolumn{1}{c}{5}   \\
\cline{1-1}
\multicolumn{1}{|c|}{4} &\multicolumn{1}{c}{6}  \\
\cline{1-1}
\end{tabular}_C
\otimes 
\begin{tabular}{|c|c|c|}
\cline{1-3} 
     1       & 2 &  \multicolumn{1}{|c|}{4} \\
\cline{1-3} 
3 & \multicolumn{1}{c}{5} & \multicolumn{1}{c}{6} \\
\cline{1-1}
\end{tabular}_{S^0}
+ 
\frac{1}{\sqrt{3}}
\begin{tabular}{|c|c|}
\hline
     1       & 2  \\
\cline{1-2} 
3  & \multicolumn{1}{c}{5}  \\
\cline{1-1}
\multicolumn{1}{|c|}{4} &\multicolumn{1}{c}{6}  \\
\cline{1-1}
\end{tabular}_C
\otimes 
\begin{tabular}{|c|c|c|}
\cline{1-3} 
     1       & 3 &  \multicolumn{1}{|c|}{4} \\
\cline{1-3} 
2 & \multicolumn{1}{c}{5} & \multicolumn{1}{c}{6} \\
\cline{1-1}
\end{tabular}_{S^0}
\Big)
\nonumber \\
-  
&\begin{tabular}{|c|c|c|}
\hline
1 & 3 &4  \\
\cline{1-3}
\multicolumn{1}{|c|}{2}  \\
\cline{1-1}
\end{tabular}_{ I}
\otimes \Big(
-\frac{1}{\sqrt{6}}
\begin{tabular}{|c|c|}
\hline
     1       & 2  \\
\cline{1-2} 
3  & \multicolumn{1}{c}{5}   \\
\cline{1-1}
\multicolumn{1}{|c|}{4} &\multicolumn{1}{c}{6}  \\
\cline{1-1}
\end{tabular}_C
\otimes 
\begin{tabular}{|c|c|c|}
\cline{1-3} 
     1       & 2 &  \multicolumn{1}{|c|}{3} \\
\cline{1-3} 
4 & \multicolumn{1}{c}{5} & \multicolumn{1}{c}{6} \\
\cline{1-1}
\end{tabular}_{S^0}
- 
\frac{1}{\sqrt{3}}
\begin{tabular}{|c|c|}
\hline
     1       & 2  \\
\cline{1-2} 
3  &  \multicolumn{1}{c}{5}   \\
\cline{1-1}
\multicolumn{1}{|c|}{4} &\multicolumn{1}{c}{6}  \\
\cline{1-1}
\end{tabular}_C
\otimes 
\begin{tabular}{|c|c|c|}
\cline{1-3} 
     1       & 2 &  \multicolumn{1}{|c|}{4} \\
\cline{1-3} 
3 & \multicolumn{1}{c}{5} & \multicolumn{1}{c}{6} \\
\cline{1-1}
\end{tabular}_{S^0}
+ 
\frac{1}{\sqrt{3}}
\begin{tabular}{|c|c|}
\hline
     1       & 3  \\
\cline{1-2} 
2  & \multicolumn{1}{c}{5}   \\
\cline{1-1}
\multicolumn{1}{|c|}{4} &\multicolumn{1}{c}{6}  \\
\cline{1-1}
\end{tabular}_C
\otimes 
\begin{tabular}{|c|c|c|}
\cline{1-3} 
     1       & 3 &  \multicolumn{1}{|c|}{4} \\
\cline{1-3} 
2 & \multicolumn{1}{c}{5} & \multicolumn{1}{c}{6} \\
\cline{1-1}
\end{tabular}_{S^0}
- 
\frac{1}{\sqrt{6}}
\begin{tabular}{|c|c|}
\hline
     1       & 4  \\
\cline{1-2} 
2  &\multicolumn{1}{c}{5}  \\
\cline{1-1}
\multicolumn{1}{|c|}{3} &\multicolumn{1}{c}{6}  \\
\cline{1-1}
\end{tabular}_C
\otimes 
\begin{tabular}{|c|c|c|}
\cline{1-3} 
     1       & 3 &  \multicolumn{1}{|c|}{4} \\
\cline{1-3} 
2 & \multicolumn{1}{c}{5} & \multicolumn{1}{c}{6} \\
\cline{1-1}
\end{tabular}_{S^0}
\Big)\Big]
\label{eq-I1}
\end{align}
\begin{align}
\vert [C, I^2, S^0]_1 \rangle=\frac{1}{\sqrt{3}}\Big[
\begin{tabular}{|c|c|c|c|}
\hline
1 & 2 & 3& 4 \\
\hline
\end{tabular}_{\quad I}
\otimes ( \quad
\begin{tabular}{|c|c|}
\hline
     1       & 2  \\
\cline{1-2} 
3  & \multicolumn{1}{c}{5}   \\
\cline{1-1}
4 &\multicolumn{1}{c}{6}  \\
\cline{1-1}
\end{tabular}_C
\otimes 
\begin{tabular}{|c|c|c|}
\cline{1-3} 
     1       & 3 & 4 \\
\cline{1-3} 
2  &   \multicolumn{1}{c}{5}  & \multicolumn{1}{c}{6} \\
\cline{1-1}
\end{tabular}_{S^0}
-\begin{tabular}{|c|c|}
\hline
     1       & 3  \\
\cline{1-2} 
2  & \multicolumn{1}{c}{5}   \\
\cline{1-1}
4 &\multicolumn{1}{c}{6}  \\
\cline{1-1}
\end{tabular}_C
\otimes 
\begin{tabular}{|c|c|c|}
\cline{1-3} 
     1       & 2 & 4 \\
\cline{1-3} 
3  &   \multicolumn{1}{c}{5}  & \multicolumn{1}{c}{6} \\
\cline{1-1}
\end{tabular}_{S^0}
+\begin{tabular}{|c|c|}
\hline
     1       & 4  \\
\cline{1-2} 
2  & \multicolumn{1}{c}{5}   \\
\cline{1-1}
3 &\multicolumn{1}{c}{6}  \\
\cline{1-1}
\end{tabular}_C
\otimes 
\begin{tabular}{|c|c|c|}
\cline{1-3} 
     1       & 2 & 3 \\
\cline{1-3} 
4 &   \multicolumn{1}{c}{5}  & \multicolumn{1}{c}{6} \\
\cline{1-1}
\end{tabular}_{S^0}
)\Big] \otimes s(5)s(6)
\label{eq-I2-1}
\end{align}
\begin{align}
\vert [C, I^2, S^0]_2 \rangle=\frac{1}{\sqrt{2}}\Big[
\begin{tabular}{|c|c|c|c|}
\hline
1 & 2 & 3& 4 \\
\hline
\end{tabular}_{\quad I}
\otimes ( \quad
\begin{tabular}{|c|c|}
\hline
     1       & 2  \\
\cline{1-2} 
3  & 4   \\
\cline{1-2}
\multicolumn{1}{c}{5} &\multicolumn{1}{c}{6}  \\
\end{tabular}_C
\otimes 
\begin{tabular}{|c|c|c|}
\cline{1-2} 
     1       & 3 &  \multicolumn{1}{c}{5} \\
\cline{1-2} 
2  & 4 & \multicolumn{1}{c}{6} \\
\cline{1-2}
\end{tabular}_{S^0}
-\begin{tabular}{|c|c|}
\hline
     1       & 3  \\
\cline{1-2} 
2  & 4   \\
\cline{1-2}
\multicolumn{1}{c}{5} &\multicolumn{1}{c}{6}  \\
\end{tabular}_C
\otimes 
\begin{tabular}{|c|c|c|}
\cline{1-2} 
     1       & 2 &  \multicolumn{1}{c}{5} \\
\cline{1-2} 
3  & 4 & \multicolumn{1}{c}{6} \\
\cline{1-2}
\end{tabular}_{S^0}
)\Big] \otimes s(5)s(6)
\label{eq-I2-2}
\end{align}
\end{widetext}

In the case for I=0, both of Eq.~(\ref{eq-F27-I0}) and Eq.~(\ref{eq-F1-I0}) can be rewritten in terms of Eq.~(\ref{eq-I0-1})
and  Eq.~(\ref{eq-I0-2}) as
\begin{align}
&\{1234\}\{56\}_{[F^{1};I^0, C, S^0]} =\nonumber \\
&-\frac{\sqrt{3}}{2}\vert [C, I^0, S^0]_1 \rangle
-\frac{1}{2}\vert [C, I^0, S^0]_2 \rangle,
\nonumber \\
&\{1234\}\{56\}_{[F^{27};I^0, C, S^0]} =\nonumber \\
&\frac{1}{2}\vert [C, I^0, S^0]_1 \rangle
-\frac{\sqrt{3}}{2}\vert [C, I^0, S^0]_2 \rangle.
\label{eq-I0-12}
\end{align}
In the case for I=2, both of Eq.~(\ref{eq-F28-I2}) and Eq.~(\ref{eq-F27-I2}) can be rewritten in terms of Eq.~(\ref{eq-I2-1})
and  Eq.~(\ref{eq-I2-2}) as 
\begin{align}
&\{1234\}\{56\}_{[F^{27};I^2, C, S^0]} =\nonumber \\
&\frac{\sqrt{2}}{\sqrt{5}}\vert [C, I^2, S^0]_1 \rangle
+\frac{\sqrt{3}}{\sqrt{5}}\vert [C, I^2, S^0]_2 \rangle,
\nonumber \\
&\{1234\}\{56\}_{[F^{28};I^2, C, S^0]} =\nonumber \\
&\frac{\sqrt{3}}{\sqrt{5}}\vert [C, I^2, S^0]_1 \rangle
-\frac{\sqrt{2}}{\sqrt{5}}\vert [C, I^2, S^0]_2 \rangle.
\label{eq-I2-12}
\end{align}

Since the states in Eq.~(\ref{eq-I0-1}), Eq.~(\ref{eq-I0-2}),  Eq.~(\ref{eq-I1}), Eq.~(\ref{eq-I2-1}), and Eq.~(\ref{eq-I2-2}) have definite symmetry properties, which are antisymmetric among particles 1, 2, 3, and 4, and at same time antisymmetric between 5 and 6, the states become eigenstates for both $-{\sum}_{i<j}^{N=4}\lambda_i^c\lambda_j^c{\sigma}_i\cdot{\sigma}_j$
and $-\lambda_5^c\lambda_6^c{\sigma}_5\cdot{\sigma}_6$, respectively. Therefore, the states are orthonormal to each other. If we focus on the Young-Yamanouchi basis, which is partly represented with a solid line in Eq.~(\ref{eq-I0-1}), Eq.~(\ref{eq-I0-2}),  Eq.~(\ref{eq-I1}), Eq.~(\ref{eq-I2-1}), and Eq.~(\ref{eq-I2-2}), the isospin, color, and spin states among particles 1, 2, 3, and 4, can be easily calculated. For example in Eq.~(\ref{eq-I0-1}), among particles 1, 2, 3, and 4, the isospin state is in a I=0 due to the Young-Yamanouchi basis [2,2], and the spin state is in a S=1 due to the Young-Yamanouchi basis [3,1], the color state is in a color singlet for three quarks and in a color triplet for one quark, while, for two quarks 5 and 6 without a solid line, the isospin state is in a I=1 due to s(5)s(6) represented by the symmetric Young-Yamanouchi basis [2], and the spin state is in a S=1 due to the same reason as the isospin state, and the color state is in a anti-triplet. In this way, the eigenstates is characterized by the following equations ;
\begin{align}
-{\sum}_{i<j}^{N=4}&\lambda_i^c\lambda_j^c{\sigma}_i\cdot{\sigma}_j \vert [C, I^0, S^0]_1 \rangle
=-\frac{16}{3} \vert [C, I^0, S^0]_1 \rangle
\nonumber \\
-&\lambda_5^c\lambda_6^c{\sigma}_5\cdot{\sigma}_6 \vert [C, I^0, S^0]_1 \rangle
=\frac{8}{3} \vert [C, I^0, S^0]_1 \rangle,
\label{eq-I0-ex1}
\end{align}
\begin{align}
-{\sum}_{i<j}^{N=4}&\lambda_i^c\lambda_j^c{\sigma}_i\cdot{\sigma}_j \vert [C, I^0, S^0]_2 \rangle
=-4 \vert [C, I^0, S^0]_2 \rangle
\nonumber \\
-&\lambda_5^c\lambda_6^c{\sigma}_5\cdot{\sigma}_6 \vert [C, I^0, S^0]_2 \rangle
=4 \vert [C, I^0, S^0]_2 \rangle,
\label{eq-I0-ex2}
\end{align}
\begin{align}
-{\sum}_{i<j}^{N=4}&\lambda_i^c\lambda_j^c{\sigma}_i\cdot{\sigma}_j \vert [C, I^1, S^0] \rangle
=\frac{8}{3} \vert [C, I^1, S^0] \rangle
\nonumber \\
-&\lambda_5^c\lambda_6^c{\sigma}_5\cdot{\sigma}_6 \vert [C, I^1, S^0] \rangle
=\frac{8}{3} \vert [C, I^1, S^0] \rangle,
\label{eq-I1-ex}
\end{align}
\begin{align}
-{\sum}_{i<j}^{N=4}&\lambda_i^c\lambda_j^c{\sigma}_i\cdot{\sigma}_j \vert [C, I^2, S^0]_1 \rangle
=\frac{56}{3} \vert [C, I^2, S^0]_1 \rangle
\nonumber \\
-&\lambda_5^c\lambda_6^c{\sigma}_5\cdot{\sigma}_6 \vert [C, I^2, S^0]_1 \rangle
=\frac{8}{3} \vert [C, I^2, S^0]_1 \rangle,
\label{eq-I2-ex1}
\end{align}
\begin{align}
-{\sum}_{i<j}^{N=4}&\lambda_i^c\lambda_j^c{\sigma}_i\cdot{\sigma}_j \vert [C, I^2, S^0]_2 \rangle
=20 \vert [C, I^2, S^0]_2 \rangle
\nonumber \\
-&\lambda_5^c\lambda_6^c{\sigma}_5\cdot{\sigma}_6 \vert [C, I^2, S^0]_2 \rangle
=4 \vert [C, I^2, S^0]_2 \rangle,
\label{eq-I2-ex2}
\end{align}
Using the above equations, Eq.~(\ref{eq-I0-12}), Eq.~(\ref{eq-I2-12}), and Eq.~(\ref{eq-I1}), the expectation value of both $-{\sum}_{i<j}^{N=4}\lambda_i^c\lambda_j^c{\sigma}_i\cdot{\sigma}_j$ and $-\lambda_5^c\lambda_6^c{\sigma}_5\cdot{\sigma}_6$ in terms of the $\{1234\}\{56\}$ state for I found in subsection 
B of section III, can be calculated, instead of directly taking the  expectation by means of the $\{1234\}\{56\}$ state. For $-\lambda_i^c\lambda_j^c{\sigma}_i\cdot{\sigma}_j$ between $i$ and $j$ ( $i$ = 1, 2, 3, 4, $j$=5, 6 ), where the number of choice for  $i$ and $j$ is 8, we note that  each of  the expectation value is the same as that of $-\lambda_1^c\lambda_5^c{\sigma}_1\cdot{\sigma}_5$, because the states have the $\{1234\}\{56\}$ symmetry. In order to calculate the expectation value of $-\lambda_1^c\lambda_5^c{\sigma}_1\cdot{\sigma}_5$, we must take advantage of the Eq.~(\ref{eq-expectation}). 

For the case of I=0 where flavor state is in a singlet before the symmetry breaking of SU(3), the expectation value of hyperfine potential in terms of a fully antisymmetric state, $\vert [F^{1},C,S^0] \rangle$ is rewritten, by using the Eq.~(\ref{eq-expectation});
\begin{align}
&\langle-{\sum}_{i<j}^{N=6}\lambda_i^c\lambda_j^c{\sigma}_i\cdot{\sigma}_j \rangle_{\vert [F^{1},C,S^0] \rangle}=
\nonumber \\
&\langle-{\sum}_{i<j}^{N=6}\lambda_i^c\lambda_j^c{\sigma}_i\cdot{\sigma}_j \rangle_{\{1234\}\{56\}_{[F^{1};I^0, C, S^0]}}=
\nonumber \\
&\langle-{\sum}_{i<j}^{N=4}\lambda_i^c\lambda_j^c{\sigma}_i\cdot{\sigma}_j \rangle_{\{1234\}\{56\}_{[F^{1};I^0, C, S^0]}}+
\nonumber \\
&8\langle -\lambda_1^c\lambda_5^c{\sigma}_1\cdot{\sigma}_5 \rangle_{\{1234\}\{56\}_{[F^{1};I^0, C, S^0]}}+
\nonumber \\
&\langle -\lambda_5^c\lambda_6^c{\sigma}_5\cdot{\sigma}_5 \rangle_{\{1234\}\{56\}_{[F^{1};I^0, C, S^0]}}.
\end{align}
Then, we have enough equations to calculate the $\langle -\lambda_1^c\lambda_5^c{\sigma}_1\cdot{\sigma}_5 \rangle_{\{1234\}\{56\}_{[F^{1};I^0, C, S^0]}}$.  Using  Eq.~(\ref{eq-I0-12}) ; 
\begin{align}
&\langle-{\sum}_{i<j}^{N=6}\lambda_i^c\lambda_j^c{\sigma}_i\cdot{\sigma}_j \rangle_{\vert [F^{1},C,S^0] \rangle}=
\nonumber \\
&\frac{3}{4}\langle-{\sum}_{i<j}^{N=4}\lambda_i^c\lambda_j^c{\sigma}_i\cdot{\sigma}_j \rangle_{\vert [C,I^0,S^0]_1 \rangle}+
\nonumber \\
&\frac{3}{4}\langle -\lambda_5^c\lambda_6^c{\sigma}_5\cdot{\sigma}_6 \rangle_{\vert [C,I^0,S^0]_1 \rangle}+
\nonumber \\
&\frac{1}{4}\langle-{\sum}_{i<j}^{N=4}\lambda_i^c\lambda_j^c{\sigma}_i\cdot{\sigma}_j \rangle_{\vert [C,I^0,S^0]_2 \rangle}+
\nonumber \\
&\frac{1}{4}\langle -\lambda_5^c\lambda_6^c{\sigma}_5\cdot{\sigma}_6 \rangle_{\vert [C,I^0,S^0]_2 \rangle}+
\nonumber \\
&8\langle -\lambda_1^c\lambda_5^c{\sigma}_1\cdot{\sigma}_5 \rangle_{[F^{1};I^0, C, S^0]}
\label{eq-expectation-1}
\end{align}
In the first line of the  Eq.~(\ref{eq-expectation-1}), the expectation value is -24, by making use of another formula~\cite{Aerts:1977rw}, which is very well known for flavor SU(3) symmetry, given by,
\begin{align}
-&{\sum}_{i<j}^{N}\lambda_i^c\lambda_j^c{\sigma}_i\cdot{\sigma}_j = \nonumber\\                                                                                                                     
[&N(N-10)+\frac{4}{3}S(S+1)+2C_c+4C_F],
\label{eq-magic-1}
\end{align}
where $C_F$=$\frac{1}{4}\lambda^F\lambda^F$. Then, we can calculate the $\langle -\lambda_1^c\lambda_5^c{\sigma}_1\cdot{\sigma}_5 \rangle_{[F^{1};I^0, C, S^0]}$, since the other terms in the Eq.~(\ref{eq-expectation-1}) are given by the Eq.~(\ref{eq-I0-ex1}) and Eq.~(\ref{eq-I0-ex2}). 
In this way, we can find the expectation value of
$-\lambda_i^c\lambda_j^c{\sigma}_i\cdot{\sigma}_j$ about the other $\{1234\}\{56\}$ state for I, without taking the trouble of         
direct calculation.

The process for calculating  the expectation value for the cross terms between different states follow a similar path. 
For the cross term between $\{1234\}\{56\}_{[F^{1};I^0, C, S^0]}$ and $\{1234\}\{56\}_{[F^{27};I^0, C, S^0]}$, the terms other than the expectation value of the $\langle -\lambda_1^c\lambda_5^c{\sigma}_1\cdot{\sigma}_5 \rangle$ are obtained by using Eq.~(\ref{eq-I0-12}), Eq.~(\ref{eq-I0-ex1}) and Eq.~(\ref{eq-I0-ex2}), while the expectation value of the $\langle -\lambda_1^c\lambda_5^c{\sigma}_1\cdot{\sigma}_5 \rangle$ is obtained by the following equation, 

\begin{align}
\langle [F^{27},C,S^0] \vert-{\sum}_{i<j}^{N=6}\lambda_i^c\lambda_j^c{\sigma}_i\cdot{\sigma}_j \vert [F^{1},C,S^0] \rangle=0.
\label{eq-expectation-cross}
\end{align}
The final results for the matrix elements are summarized in  Table \ref{hyperfine-matrix}.  The corresponding matrix elements for the baryon in Isospin symmetric states are given in Table \ref{hyperfine-matrix-baryon}.

\begin{table}[htdp]
\caption{The matrix element of $-\langle \lambda_i^c\lambda_j^c{\sigma}_i\cdot{\sigma}_j \rangle$ for hyperfine potential of the dibaryon with respect to isospin and flavor.  }
\begin{center}
\begin{tabular}{c|c|c|c}
\hline \hline
  Isospin,               & $-\langle \lambda_i^c\lambda_j^c{\sigma}_i\cdot{\sigma}_j \rangle$   & $-\langle \lambda_i^c\lambda_j^c{\sigma}_i\cdot{\sigma}_j \rangle$      &   $-\langle \lambda_i^c\lambda_j^c{\sigma}_i\cdot{\sigma}_j \rangle$                        \\
  Flavor              & $i$$<$$j$ =1$\sim$4      &  $i$=1$\sim$4,    $j$=5, 6         & $i$=5, $j$=6       \\
\hline 
 I=0,   $F^1$                 &        -5/6                    &        -11/4                           &            3          \\ 
I=0,   $F^{27}$               &        -13/18                &        13/12                           &           11/3          \\
 Cross terms                   &  1/(6$\sqrt{3}$)         &          -1/(4$\sqrt{3}$)               &     1/$\sqrt{3}$   \\
\hline 
 I=1,   $F^{27}$               &        4/9                    &        1/3                          &        8/3          \\ 
\hline 
 I=2,   $F^{28}$                 &        16/5                    &        16/5                           &           16/5          \\ 
I=2,   $F^{27}$               &        146/45                &        -28/15                          &           52/15          \\
 Cross terms                   &  -2$\sqrt{2}$/(15$\sqrt{3}$)   &   $\sqrt{2}$/(5$\sqrt{3}$)   &  -4$\sqrt{2}$/(5$\sqrt{3}$)      \\
\hline \hline
\end{tabular}
\end{center}
\label{hyperfine-matrix}
\end{table}

\begin{table}[htdp]
\caption{The matrix element of $-\langle \lambda_i^c\lambda_j^c{\sigma}_i\cdot{\sigma}_j \rangle$ for hyperfine potential of the baryon spectrum.  }
\begin{center}
\begin{tabular}{c|c|c|c}
\hline \hline
  Baryon            & $-\langle \lambda_1^c\lambda_2^c{\sigma}_1\cdot{\sigma}_2 \rangle$   & $-\langle \lambda_1^c\lambda_3^c{\sigma}_1\cdot{\sigma}_3 \rangle$      &   $-\langle \lambda_2^c\lambda_3^c{\sigma}_2\cdot{\sigma}_3 \rangle$                        \\
 \hline 
      N           &        -8/3                   &        -8/3                           &            -8/3          \\ 
 \hline       
  $\Lambda$      &        -8                &        0                           &           0        \\
   \hline   
  $\Xi$             &  8/3              &          -16/3               &     -16/3   \\
\hline 
$\Sigma$              &        8/3                   &        -16/3                          &        -16/3          \\ 
\hline \hline 
\end{tabular}
\end{center}
\label{hyperfine-matrix-baryon}
\end{table}

It is interesting to discuss the hyperfine factors for the difference between a dibaryon state and its lowest two baryon threshold.  In the SU(3) symmetric limit, all quarks will have the same mass.  In this case, the hyperfine factors appearing in the three columns in Tables \ref{hyperfine-matrix} and \ref{hyperfine-matrix-baryon} will contribute with equal strength to the potential.  Therefore, the factor for $I=0,F^1$ state is $6 \times (-5/6)+8\times (-11/4) +3 =-24$, while that for the two $\Lambda$ state is -16, resulting in an attractive contribution with a factor -8.  However, consider the limit where the strange quark mass becomes infinitely heavy.  Then, the contribution from the second and third column in both Table \ref{hyperfine-matrix} and \ref{hyperfine-matrix-baryon} will not contribute.  This will not have any effect on the two $\Lambda$ threshold but the dibaryon hyperfine factor will now become just -5.  Hence, it is natural that the dibaryon could become stable when the strange quark mass decreases.  However, when the SU(3) symmetric mass increases, the attraction will become smaller.  As will be borne out in the next section, although one still can not find a bound state, the general tendency seems to be true.

For the expectation value of $\lambda_i^c\lambda_j^c$ appearing in the confinement potential, the matrix elements are calculated in the same procedure as in the  hyperpotential case, and using the followings;

\begin{align}
{\sum}_{i<j}^{N=6}\lambda_i^c\lambda_j^c=-\frac{8}{3}N+2C_c.
\label{eq-lambda}
\end{align}

\begin{table}[htdp]
\caption{The matrix element of $-\langle \lambda_i^c\lambda_j^c \rangle$ for confinement potential with respect to isospin and flavor.  }
\begin{center}
\begin{tabular}{c|c|c|c}
\hline \hline
  Isospin,               & $-\langle \lambda_i^c\lambda_j^c \rangle$   & $-\langle \lambda_i^c\lambda_j^c \rangle$      &   $-\langle \lambda_i^c\lambda_j^c \rangle$                        \\
  Flavor              & $i$$<$$j$ =1$\sim$4      &  $i$=1$\sim$4,    $j$=5, 6         & $i$=5, $j$=6       \\
\hline 
 I=0,   $F^1$                 &        7/6                    &        11/12                           &           5/3          \\ 
I=0,   $F^{27}$               &        5/6                &        17/12                           &           -1/3          \\
 Cross terms                   &  1/(2$\sqrt{3}$)         &          -$\sqrt{3}$/4               &     $\sqrt{3}$   \\
\hline 
 I=1,   $F^{27}$               &        4/3                    &        2/3                          &        8/3          \\ 
\hline 
 I=2,   $F^{28}$                 &        16/15                    &        16/15                           &           16/15          \\ 
I=2,   $F^{27}$               &        14/15                &        19/15                          &           4/15          \\
 Cross terms                   &  -2$\sqrt{2}$/(5$\sqrt{3}$)   &   $\sqrt{6}$/5   &  -4$\sqrt{6}$/5     \\
\hline \hline
\end{tabular}
\end{center}
\label{confinement-matrix}
\end{table}

We summarize all matrix elements for both hyperfine and confinement potential of the dibaryon   in Table~\ref{hyperfine-matrix} and Table~\ref{confinement-matrix}. To compare  matrix elements for  hyperfine potential of the dibaryon, we show that of the
baryon in  Table~\ref{hyperfine-matrix-baryon}. For matrix elements for confinement potential of the baryon, it is very  easy to
show from the Eq.~(\ref{eq-lambda}) that $-\langle \lambda_i^c\lambda_j^c \rangle$ ( $i$$<$$j$ =1, 2, 3 ) is 8/3.
We note that direct calculations of the expectation value of both hyperfine and confinement potential in terms of  $\{1234\}\{56\}$ state, which is obtained in subsection B of section III, is equal to those mentioned above.

\section{Numerical Results}  

In this section, we analyze the numerical results obtained from the variational method, by using a total wave function that 
consists of the spatial function in Eq.~(\ref{eq-spatial}) as the trial function and the color $\otimes$ isospin $\otimes$ spin state constructed in the previous section. Since there are two color $\otimes$ isospin $\otimes$ spin states for I=0, given by $\{1234\}\{56\}_{[F^{1};I^0, C, S^0]}$ and $\{1234\}\{56\}_{[F^{27};I^0, C, S^0]}$, the expectation value of the Hamiltonian is a 2 by 2 matrix in terms of the two color $\otimes$ isospin $\otimes$ spin states. Therefore,
the ground state for I=0 must be represented as the mixing form of $\{1234\}\{56\}_{[F^{1};I^0, C, S^0]}$ and $\{1234\}\{56\}_{[F^{27};I^0, C, S^0]}$. For the same reason, the ground state for I=2 must be represented as a mixed state of $\{1234\}\{56\}_{[F^{27};I^2, C, S^0]}$ and $\{1234\}\{56\}_{[F^{28};I^2, C, S^0]}$. 

Table~\ref{dibaryon-mass} shows the result of the analysis for the mass of the dibaryons containing two strange quarks with S=0, with respect to I. Table~\ref{dibaryon-matrix-element} shows the matrix element of the expectation value of the Hamiltonian for I=0 and I=2. As we see in Table~\ref{dibaryon-mass}, there are no bound dibaryons againt the strong decay into two baryons. 
It should be emphasized that our approach only probes compact six quark states based on color confining and color-spin potential.  To probe molecular configuration, we have to include possible long range forces induced by meson exchange potentials. Moreover, the trial spatial wave function should have sufficient parameters to allow for largely separated two baryon states.  
 Therefore, the nonexistence proves that there are no compact six quark dibaryon states possible.  In I=0 and I=2, the $\{1234\}\{56\}_{[F^{1};I^0, C, S^0]}$ state and  $\{1234\}\{56\}_{[F^{27};I^2, C, S^0]}$ state are overwhelmingly dominant terms in each of ground state, respectively, so that the mixing effect is nearly negligible.
\begin{table}[htdp]
\caption{The mass of the dibaryon with respect to (I,S=0) state. The binding energy $E_B$ is taken to be the difference between the mass of the dibaryon and the lowest two baryon threshold. The unit of the mass and the variational parameters are  $\rm{GeV}$, and fm$^{-2}$, respectively. }
\begin{center}
\begin{tabular}{c|c|c|c}
\hline \hline
    (I,S)                &  (0,0)          &  (1,0)         & (2,0)         \\
\hline
Type                  &  $uuddss$    &  $uuudss$  & $uuuuss$ \\
\hline
                      
 Mass               &   2.549     &  2.88     &  2.918           \\
\hline
Variational           & $a$=2.3,        &  $a$=1.9,     & $a$=1.5        \\
parameters          &  $b$=1.7,       & $b$=1.1,      & $b$=0.9,         \\
                          &  $c$=2.1        & $c$=1.1       & $c$=1.7          \\
\hline
Decay mode     &  $\Lambda \Lambda$ & $\Xi$N  & $\Sigma \Sigma$                     \\
\hline
 $E_B$             & 0.319      & 0.562     & 0.484                \\
\hline \hline
\end{tabular}
\end{center}
\label{dibaryon-mass}
\end{table}
\begin{table}[htdp]
\caption{The matrix for the expectation values of the Hamiltonian.  The upper and lower matrices are for I=0 and I=2, respectively.}
\begin{center}
\begin{tabular}{c|c}
\hline \hline
 Basis functions   & $uuddss$  \\
\hline  
 
 $[F^{27};I^0, C, S^0]$, $[F^{1};I^0, C, S^0]$ & $\left(\begin{array}{ccccc} 2.967   & -0.0018    \\ -0.0018  & 2.549 \end{array} \right)$ \\
\hline 
  Basis functions  & $uuuuss$  \\
\hline  
  $[F^{28};I^2, C, S^0]$, $[F^{27};I^2, C, S^0]$            & $\left(\begin{array}{ccccc} 3.219   & -0.0168    \\ -0.0168  & 2.919  \end{array} \right)$ \\
\hline \hline             
\end{tabular}
\end{center}
\label{dibaryon-matrix-element}
\end{table}

So far, we investigated the stability of the dibaryon in the realistic case of broken  SU(3) falvor, where we took $m_s$ to be  heavier than  $m_u$ (=$m_d$). We now consider the possibility of stable H-dibaryon in the flavor SU(3) symmetric limit. Such possibility was recently  proposed by lattice calculation~\cite{Beane:2010hg,Inoue:2010es}. 
In Ref.~\cite{Inoue:2010es}, using the baryon-baryon potential  extracted from lattice QCD in flavor SU(3) limit, the H-dibaryon was found to be stable               
for pseudoscalar meson mass of 673-1015 $\rm{MeV}$.  Under such  circumstance, since the strange quark mass $m_s$ will be  identified with the $m_u$, the flavor $\otimes$ color $\otimes$ spin state should be fully antisymmetric, only if we choose the spatial function to be fully symmetric. So, 
for the flavor $\otimes$ color $\otimes$ spin state with full antisymmetry, we use the $\vert [F^{1},C,S^0] \rangle$ state in Eq.~(\ref{eq-F1-C2}), and for the spatial function with full symmetry, we use that introduced in the our previous paper~\cite{Park:2015nha}. 
Moreover, to compare our work with the result of the previous paper, we try to  search for the existence of the stable H-dibaryon in a wide range of $\pi$ meson mass.                      
                                  
For this purpose, we will keep most of the parameters in Eq.~(\ref{eq-hamiltonian}) the same, but refit D and ${\kappa}_0$ to better reproduce the meson spectrum.  
The fitting parameters, including $m_c$ and $m_b$, are given in Table~\ref{fitting-parameter-2}.  The meson  masses obtained from the variational method with the fitting parameters are given in Table~\ref{fitting-mass-2}.

\begin{table}[htdp]
\caption{ Parameters fitted to the experimental meson masses. }
\begin{center}
\begin{tabular}{c|c|c|c|c|c}
\hline \hline
Parameter & D & ${\kappa}_0$ &   $m_u$    & $m_c$  & $m_b$    \\
\hline
  Value  & 0.954     &  0.252  & 0.347 &  1.793 &  5.23 \\
\hline 
 Unit & $\rm{GeV}$  &$\rm{GeV}$  &  $\rm{GeV}$   &   $\rm{GeV}$ &   $\rm{GeV}$ \\ 
\hline \hline  
\end{tabular}
\end{center}
\label{fitting-parameter-2}
\end{table}
\begin{table}[htdp]
\caption{Meson masses obtained from the variational method. The third row indicates the experimental data~\cite{Agashe:2014kda}. (units are $\rm{GeV}$ ) }
\begin{center}
\begin{tabular}{c|c|c|c|c|c|c|c|c}
\hline \hline
Meson     & $\pi$     & $\rho$  & $D$ & $D^*$    & ${\eta}_c$    & J/$\psi$     & $\Upsilon$    &  ${\eta}_b$         \\
\hline
Mass      & 0.146          & 0.775   & 1.892   & 2.024   & 2.989       & 3.096       & 9.398       & 9.471       \\                     
\hline
Exp       &  0.139   &  0.775  & 1.869    &  2.006  &  2.983            & 3.096    & 9.398      & 9.46      \\
            \hline \hline
\end{tabular}
\end{center}
\label{fitting-mass-2}
\end{table}

\begin{figure}[htbp]
  \begin{center}
\includegraphics[scale=0.8]{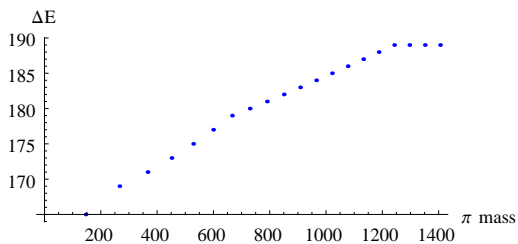}
\caption[]{The the mass difference ($\Delta E$) between the H-dibaryon and two $\Lambda$ baryons as a function of the pion mass in SU(3) limit.  (units are $\rm{MeV}$ ) }
\label{fig-1}
  \end{center}
\end{figure}

\begin{figure}[htbp]
  \begin{center}
\includegraphics[scale=0.9]{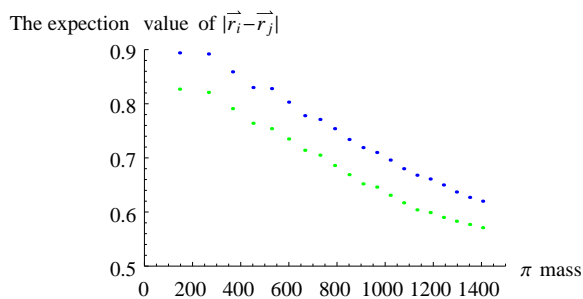}
\caption[]{ The expectation value of the distance between interquarks ($\langle \mid\textbf{r}_i-\textbf{r}_j\mid \rangle$) for the H-dibaryon (blue dotted curve) and $\Lambda$ baryon (green dotted curve) as a function of the pion mass in SU(3) limit. The  expectation value of $\mid\textbf{r}_i-\textbf{r}_j\mid$ are the same for any $i$ and $j$ ($i$ $<$ $j$ ) as the spatial wave function for both the H-dibaryon and $\Lambda$ baryon are fully symmetric. (units: $x-$axis MeV, $y-$axis fm) }
\label{fig-2}
  \end{center}
\end{figure}

We now increase $\pi$ mass smoothly by varying the constituent quark mass from 347 $\rm{MeV}$ to 917 $\rm{MeV}$. The resulting mass difference between the H-dibaryon and two $\Lambda$ bayrons are plotted in Fig.~\ref{fig-1} as a function of the pion mass.  First, it should be noted that when $m_s$ is reduced to $m_u$, the mass difference  decreases to below 170 MeV, which is smaller than that of the realistic case given in Table VI.  This result is consistent with the hyperfine interaction becoming larger as discussed in the previous section.  On the other hand, as one increases the pion mass, as can be seen in Fig.~\ref{fig-1}, it is found that the mass difference between the H-dibaryon and two $\Lambda$ baryons  increases monotonously. This seems to be a consequence of the overall weakening of the hyperfine interaction due to the prefactors proportional to the inverse quark masses. The wave function is also becoming more compact as can be seen from Fig.\ref{fig-2} which makes the contribution from the additional kinetic term larger.    Hence, one can conclude that even in the flavor SU(3) symmetric limit, there is no stable compact H-dibaryon for a wide range of pion mass.  But to probe a possible molecular bound state, further input in the Hamiltonian and trial wave function would be needed.  
\section{Summary}

For the dibaryon containing two strange quarks with S=0, such as, $uuddss$, $uuudss$, and $uuuuss$, we have constructed the flavor $\otimes$ color $\otimes$ spin state both in the flavor SU(3) symmetric and breaking cases appropriate for the symmetric spatial wave function,  satisfying the Pauli principle. We showed that the possible dibaryon states could be classified in terms of the symmetry property of Young tableau. It is found that
the symmetry breaking of flavor SU(3) reduces the fully antisymmetric states in flavor SU(3) into  the color $\otimes$ isospin $\otimes$ spin states in flavor SU(2). 
 
In order to investigate the stability of the dibaryon, we adopted the nonrelativistic Hamiltonian, including confinement and hyperfine potential, and calculated the mass of the dibaryon, by using the variational method.   
We conclude that there are no compact bound dibaryons states that are stable against the strong decay into two baryons in both the symmetric and symmetry breaking limit of flavor SU(3).                                                                     
Further improvements in the potential as well as more sophisticated spatial trial wave function are needed to probe largely separated two baryon molecular bound state.

\textit{Acknowledgements} This work was supported by the Korea
National Research Foundation under the grant number
KRF-2011-0020333 and KRF-2011-0030621.                               
The work of W.S. Park is supported in part by the Yonsei University  Research Fund(Post Doc. Researcher Supporting Program) of 2014 (project no.: 2014-12-0139).

\appendix

\section{Flavor state in SU(3)}                                   
 
In this Appendix A, we systematically investigate how to find out the flavor state, in which lie SU(3), by introducing a generator operator. As mensioned in subsection C of section III, we presented the flavor states, which correspond to Young-Yamanouchi basis. In dealing with each of flavor multiplet, in particular, it shoud be noted that for a given flavor multiplet, the representation of the
flavor states are different with respect to isospin. First, we consider flavor 27-multiplet with 9 dimension, concerning I=0, I=1, and I=2. For given I, once we find one state among the flavor multiplet, the others can be easily obtained by applying a permutation operator.

\subsection{ Flavor 27-multiplet with I=0} 

The $\vert F_1^{27} \rangle$ state with I=0 is obtained by introducing the following generator operator;                                                       
\begin{align}                                                                 
G_{1,I=0}^{27} \equiv \frac{1}{\sqrt{12}}\frac{1}{2!}\frac{1}{2!}{\sum}_{\sigma \in S_4}\sigma A_{1,I=0}^{27},
\label{eq-generator-F27-I0}
\end{align}                                                                 
where $A_{1,I=0}^{27}$=$(A_{15}A_{26}+A_{25}A_{16})$, $A_{ij}$=1 - ($ij$), and $S_4$ is permutation group for particles 1, 2, 3, and 4. \\
The action of the $A_{1,I=0}^{27}$ on $u(1)u(2)s(3)s(4)d(5)d(6)$ gives $2uussdd$ + $2ddssuu$ - $udssud$ - $udssdu$ - $dussud$ - $dussdu$, which will be denoted by $I^0_1s(3)s(4)$, because the four quarks except for two strange quarks s(3) and s(4) is in the $I^0_1$ represented in Eq.~(\ref{eq-I-0}). Since $\pi {\sum}_{\sigma \in S_4}\sigma$ = ${\sum}_{\sigma \in S_4}\sigma$ for $\pi \in S_4$ and (56)$A_{1,I=0}^{27}$=$A_{1,I=0}^{27}$(56), the $G_{1,I=0}^{27}$ $u(1)u(2)s(3)s(4)d(5)d(6)$ represent the $\vert F_1^{27} \rangle$ state with I=0, whose the symmetry property has [1234][56].

\subsection{ Flavor 27-multiplet with I=1} 

The $\vert F_1^{27} \rangle$ state with I=1 is obtained by introducing the following generator operator;                                                       
\begin{align}                                                                 
G_{1,I=1}^{27} \equiv \frac{1}{8\sqrt{3}}\frac{1}{3!}{\sum}_{\sigma \in S_4}\sigma S_{56} A_{1,I=1}^{27},
\label{eq-generator-F27-I1}
\end{align}                                                                 
where $A_{1,I=1}^{27}$=$(A_{15}A_{26}+A_{25}A_{36}+A_{35}A_{16})$, and $S_{56}$=1 + (56). \\
 The action of the $G_{1,I=1}^{27}$ on $u(1)u(2)u(3)s(4)d(5)s(6)$ gives the $\vert F_1^{27} \rangle$ state with I=1, whose the symmetry property has [1234][56], for the same reason.

\subsection{ Flavor 27-multiplet with I=2} 

The $\vert F_1^{27} \rangle$ state with I=2 is obtained by introducing the following generator operator;                                                       
\begin{align}                                                                 
G_{1,I=2}^{27} \equiv \frac{1}{4\sqrt{15}}\frac{1}{2!}\frac{1}{2!}{\sum}_{\sigma \in S_4}\sigma A_{1,I=2}^{27},
\label{eq-generator-F27-I2}
\end{align}                                                                 
where $A_{1,I=2}^{27}$=$(A_{15}A_{26}+A_{25}A_{16})$. \\
The action of the $A_{1,I=2}^{27}$ on $u(1)u(2)u(3)u(4)s(5)s(6)$ gives $2uuuuss$ + $2ssuuuu$ - $suuuus$ - $suuusu$ - $usuusu$ - $usuuus$, which will be denoted by 2$I^2s(5)s(6)$ + 2$I^2s(1)s(2)$ - $I^2s(1)s(6)$ - $I^2s(1)s(5)$ - $I^2s(2)s(5)$ - $I^2s(2)s(6)$, because the four quarks except for two strange quarks is in the $I^2$ represented in Eq.~(\ref{eq-I-2}). Then, $G_{1,I=2}^{27}$ $u(1)u(2)u(3)u(4)s(5)s(6)$ represent the $\vert F_1^{27} \rangle$ state with I=2, whose the symmetry property has [1234][56], for the same reason.

\subsection{ Flavor 28-multiplet with I=2} 
For $\vert F^{28} \rangle$ state with 1 dimension, the state with I=2 is obtained by introducing the following generator operator;                                                       
\begin{align}                                                                 
G_{I=2}^{28} \equiv \frac{1}{\sqrt{15}}\frac{1}{4!}\frac{1}{2!}{\sum}_{\sigma \in S_6}\sigma,
\label{eq-generator-F28-I2}
\end{align}                                                                 
where $S_6$ is permutation group for particles 1, 2, 3, 4, 5, and 6. \\
The action of the $G_{I=2}^{28}$ on $u(1)u(2)u(3)u(4)s(5)s(6)$ gives the $\vert F^{28} \rangle$ state with I=2, whose the symmetry property has [123456], because $\pi {\sum}_{\sigma \in S_6}\sigma$ = ${\sum}_{\sigma \in S_6}\sigma$ for $\pi \in S_6$.

\subsection{ Flavor singlet with I=0} 

For flavor singlet, it is convenient to deal with $\vert F^{1}_5 \rangle$ state, because the flavor singlet state 
come from the flavor singlet for the particles 1, 2, and 3, and the flavor singlet for the particles 4, 5, and 6.
The $\vert F^{1}_5 \rangle$ state with I=0 is obtained by introducing the following generator operator ;                                                       
\begin{align}                                                                 
G_{5,I=0}^{1} \equiv \frac{1}{6}{\sum}_{\sigma \in S_3}{(-1)}^{\sigma}\sigma{\sum}_{\pi \in S_3}{(-1)}^{\pi}\pi,
\label{eq-generator-F1-I0}
\end{align}                                                                 
where ${(-1)}^{\sigma}$ is 1 if $\sigma$ is an even permutation, and -1 if $\sigma$ is an odd permutation, and the first permutation group $S_3$ is applied on the particles 1, 2, and 3, and the sencond, the particles 4, 5, and 6. 
The action of the $G_{5,I=0}^{1}$ on $u(1)d(2)s(3)u(4)d(5)s(6)$ gives the $\vert F^{1}_5 \rangle$ state with I=0, whose the symmetry property has $\{123\}\{456\}$,  because $\pi {\sum}_{\sigma \in S_3}{(-1)}^{\sigma}\sigma$ = ${(-1)}^{\pi} {\sum}_{\sigma \in S_3}{(-1)}^{\sigma}\sigma$ for $\pi \in S_3$.

\begin{widetext}\allowdisplaybreaks
$F^{27}$ multiplet for I=0; \\
\begin{align}
\vert F^{27}_1 \rangle=\frac{1}{6\sqrt{2}}[I^0_1s(1)s(2)+I^0_1s(1)s(3)+I^0_1s(1)s(4)+
I^0_1s(2)s(3)+I^0_1s(2)s(4)+I^0_1s(3)s(4)],\nonumber
\end{align}
\begin{align}
\vert F^{27}_2 \rangle=\frac{\sqrt{2}}{3\sqrt{15}}[&-\frac{1}{4}I^0_1s(1)s(2)+\frac{3}{2}I^0_2s(1)s(2)-\frac{1}{4}I^0_1s(1)s(3)+\frac{3}{2}I^0_2s(1)s(3)+I^0_1s(1)s(5)+\frac{1}{4}I^0_1s(1)s(4)-\frac{1}{4}I^0_1s(2)s(3) \nonumber \\
&+\frac{3}{2}I^0_2s(2)s(3)+I^0_1s(2)s(5)+\frac{1}{4}I^0_1s(2)s(4)+I^0_1s(3)s(5)+\frac{1}{4}I^0_1s(3)s(4)], \nonumber
\end{align}
\begin{align}
\vert F^{27}_3 \rangle=\frac{1}{2\sqrt{15}}[&-\frac{1}{3}I^0_1s(1)s(2)-I^0_2s(1)s(2)-\frac{1}{6}I^0_1s(1)s(4)+\frac{1}{6}I^0_1s(1)s(3)+\frac{3}{2}I^0_2s(1)s(4)+\frac{1}{2}I^0_2s(1)s(3)-\frac{1}{6}I^0_1s(1)s(5) \nonumber \\
&+\frac{3}{2}I^0_2s(1)s(5)-\frac{1}{6}I^0_1s(2)s(4)+\frac{1}{6}I^0_1s(2)s(3)+\frac{3}{2}I^0_2s(2)s(4)+\frac{1}{2}I^0_2s(2)s(3)-\frac{1}{6}I^0_1s(2)s(5)+\frac{3}{2}I^0_2s(2)s(5)\nonumber \\
&+I^0_1s(4)s(5)+\frac{1}{3}I^0_1s(3)s(5)+\frac{1}{3}I^0_1s(3)s(4)], \nonumber
\end{align}
\begin{align}
\vert F^{27}_4 \rangle=\frac{1}{\sqrt{45}}[&-\frac{1}{4}I^0_1s(1)s(3)-\frac{3}{4}I^0_2s(1)s(3)-\frac{1}{4}I^0_1s(1)s(4)-\frac{3}{4}I^0_2s(1)s(4)-\frac{1}{4}I^0_1s(1)s(5)-\frac{3}{4}I^0_2s(1)s(5)+\frac{1}{4}I^0_1s(2)s(4) \nonumber \\
&+\frac{1}{4}I^0_1s(2)s(3)+\frac{3}{2}I^0_2s(3)s(4)+\frac{3}{4}I^0_2s(2)s(4)+\frac{3}{4}I^0_2s(2)s(3)+\frac{1}{4}I^0_1s(2)s(5)+\frac{3}{2}I^0_2s(3)s(5)+\frac{3}{4}I^0_2s(2)s(5)\nonumber \\
&+\frac{3}{2}I^0_2s(4)s(5)], \nonumber
\end{align}
\begin{align}
\vert F^{27}_5 \rangle=\frac{1}{2\sqrt{15}}[&-\frac{1}{3}I^0_1s(1)s(2)-I^0_2s(1)s(2)-\frac{1}{3}I^0_1s(1)s(3)-I^0_2s(1)s(3)+I^0_1s(1)s(6)+\frac{1}{3}I^0_1s(1)s(5)+\frac{1}{3}I^0_1s(1)s(4) \nonumber \\
&-\frac{1}{3}I^0_1s(2)s(3)-I^0_2s(2)s(3)+I^0_1s(2)s(6)+\frac{1}{3}I^0_1s(2)s(5)+\frac{1}{3}I^0_1s(2)s(4)+I^0_1s(3)s(6)+\frac{1}{3}I^0_1s(3)s(5)\nonumber \\
&+\frac{1}{3}I^0_1s(3)s(4)], \nonumber
\end{align}
\begin{align}
\vert F^{27}_6 \rangle=\frac{\sqrt{3}}{4\sqrt{10}}[&-\frac{4}{9}I^0_1s(1)s(2)+\frac{2}{3}I^0_2s(1)s(2)-\frac{2}{9}I^0_1s(1)s(4)+\frac{2}{9}I^0_1s(1)s(3)-I^0_2s(1)s(4)-\frac{1}{3}I^0_2s(1)s(3)-\frac{1}{6}I^0_1s(1)s(6) \nonumber \\
&+\frac{3}{2}I^0_2s(1)s(6)-\frac{1}{18}I^0_1s(1)s(5)+\frac{1}{2}I^0_2s(1)s(5)-\frac{2}{9}I^0_1s(2)s(4)+\frac{2}{9}I^0_1s(2)s(3)-I^0_2s(2)s(4)-\frac{1}{3}I^0_2s(2)s(3)\nonumber \\
&-\frac{1}{6}I^0_1s(2)s(6)+\frac{3}{2}I^0_2s(2)s(6)-\frac{1}{18}I^0_1s(2)s(5)+\frac{1}{2}I^0_2s(2)s(5)+I^0_1s(4)s(6)+\frac{1}{3}I^0_1s(3)s(6)+\frac{1}{3}I^0_1s(4)s(5) \nonumber \\
&+\frac{1}{9}I^0_1s(3)s(5)+\frac{4}{9}I^0_1s(3)s(4)
], \nonumber
\end{align}
\begin{align}
\vert F^{27}_7 \rangle=\frac{1}{2\sqrt{10}}[&-\frac{1}{3}I^0_1s(1)s(3)+\frac{1}{2}I^0_2s(1)s(3)-\frac{1}{3}I^0_1s(1)s(4)+\frac{1}{2}I^0_2s(1)s(4)-\frac{1}{4}I^0_1s(1)s(6)-\frac{3}{4}I^0_2s(1)s(6)-\frac{1}{12}I^0_1s(1)s(5) \nonumber \\
&-\frac{1}{4}I^0_2s(1)s(5)+\frac{1}{3}I^0_1s(2)s(4)+\frac{1}{3}I^0_1s(2)s(3)-I^0_2s(3)s(4)-\frac{1}{2}I^0_2s(2)s(4)-\frac{1}{2}I^0_2s(2)s(3)+\frac{1}{4}I^0_1s(2)s(6)\nonumber \\
&+\frac{3}{2}I^0_2s(3)s(6)+\frac{3}{4}I^0_2s(2)s(6)+\frac{1}{12}I^0_1s(2)s(5)+\frac{1}{2}I^0_2s(3)s(5)+\frac{1}{4}I^0_2s(2)s(5)+\frac{3}{2}I^0_2s(4)s(6)+\frac{1}{2}I^0_2s(4)s(5) \nonumber \\
&], \nonumber
\end{align}
\begin{align}
\vert F^{27}_8 \rangle=\frac{1}{4\sqrt{10}}[&-\frac{2}{3}I^0_1s(1)s(2)-\frac{1}{3}I^0_1s(1)s(3)-\frac{1}{3}I^0_1s(1)s(4)-\frac{1}{2}I^0_1s(1)s(5)-\frac{3}{2}I^0_2s(1)s(5)-\frac{1}{2}I^0_1s(1)s(6)-\frac{3}{2}I^0_2s(1)s(6) \nonumber \\
&-\frac{1}{3}I^0_1s(2)s(3)-\frac{1}{3}I^0_1s(2)s(4)-\frac{1}{2}I^0_1s(2)s(5)-\frac{3}{2}I^0_2s(2)s(5)-\frac{1}{2}I^0_1s(2)s(6)-\frac{3}{2}I^0_2s(2)s(6)+\frac{2}{3}I^0_1s(3)s(4)\nonumber \\
&+I^0_1s(3)s(5)+I^0_1s(3)s(6)+I^0_1s(4)s(5)+I^0_1s(4)s(6)+2I^0_1s(5)s(6)], \nonumber
\end{align}
\begin{align}
\vert F^{27}_9 \rangle=\frac{\sqrt{3}}{4\sqrt{10}}[&\frac{1}{3}I^0_1s(1)s(3)-\frac{1}{3}I^0_1s(1)s(4)-\frac{1}{2}I^0_1s(1)s(5)+\frac{1}{2}I^0_2s(1)s(5)-\frac{1}{2}I^0_1s(1)s(6)+\frac{1}{2}I^0_2s(1)s(6)-\frac{1}{3}I^0_1s(2)s(3)+ \nonumber \\
&\frac{1}{3}I^0_1s(2)s(4)+\frac{1}{2}I^0_1s(2)s(5)-\frac{1}{2}I^0_2s(2)s(5)+\frac{1}{2}I^0_1s(2)s(6)-\frac{1}{2}I^0_2s(2)s(6)-I^0_2s(3)s(5)-I^0_2s(3)s(6)+\nonumber \\
&I^0_2s(4)s(5)+I^0_2s(4)s(6)+2I^0_2s(5)s(6)]. 
\end{align}
$F^{27}$ multiplet for I=1; \\
\begin{align}
\vert F^{27}_1 \rangle=&-\frac{1}{12\sqrt{3}}I^1_1s(1)s(2)-\frac{1}{12\sqrt{3}}I^1_1s(1)s(3)-\frac{1}{12\sqrt{3}}I^1_1s(1)s(4)+\frac{1}{8\sqrt{3}}I^1_1s(1)s(5)+\frac{1}{8\sqrt{3}}I^1_1s(1)s(6)-\frac{1}{12\sqrt{3}}I^1_1s(2)s(3)\nonumber \\
&-\frac{1}{12\sqrt{3}}I^1_1s(2)s(4) +\frac{1}{8\sqrt{3}}I^1_1s(2)s(5)+\frac{1}{8\sqrt{3}}I^1_1s(2)s(6)-\frac{1}{12\sqrt{3}}I^1_1s(3)s(4)+\frac{1}{8\sqrt{3}}I^1_1s(3)s(5)+\frac{1}{8\sqrt{3}}I^1_1s(3)s(6)\nonumber \\
&+\frac{1}{8\sqrt{3}}I^1_1s(4)s(5)+\frac{1}{8\sqrt{3}}I^1_1s(4)s(6)
-\frac{1}{6\sqrt{3}}I^1_2s(1)s(2)-\frac{1}{6\sqrt{3}}I^1_2s(1)s(3)-\frac{1}{6\sqrt{3}}I^1_2s(1)s(4)-\frac{1}{6\sqrt{3}}I^1_2s(2)s(3) \nonumber \\
&-\frac{1}{6\sqrt{3}}I^1_2s(2)s(4)-\frac{1}{6\sqrt{3}}I^1_2s(3)s(4)], \nonumber
\end{align}
\begin{align}
\vert F^{27}_2 \rangle=&-\frac{\sqrt{5}}{36}I^1_1s(1)s(2)-\frac{\sqrt{5}}{36}I^1_1s(1)s(3)+\frac{\sqrt{5}}{36}I^1_1s(1)s(4)-\frac{\sqrt{5}}{72}I^1_1s(1)s(5)-\frac{1}{72\sqrt{5}}I^1_1s(1)s(6)-\frac{\sqrt{5}}{36}I^1_1s(2)s(3)\nonumber \\
&+\frac{\sqrt{5}}{36}I^1_1s(2)s(4)-\frac{\sqrt{5}}{72}I^1_1s(2)s(5)-\frac{1}{72\sqrt{5}}I^1_1s(2)s(6)+\frac{\sqrt{5}}{36}I^1_1s(3)s(4)-\frac{\sqrt{5}}{72}I^1_1s(3)s(5)-\frac{1}{72\sqrt{5}}I^1_1s(3)s(6)\nonumber \\
&+\frac{\sqrt{5}}{24}I^1_1s(4)s(5)+\frac{1}{24\sqrt{5}}I^1_1s(4)s(6)
+\frac{1}{6\sqrt{5}}I^1_1s(5)s(6)+\frac{1}{18\sqrt{5}}I^1_2s(1)s(2)+\frac{1}{18\sqrt{5}}I^1_2s(1)s(3)-\frac{1}{18\sqrt{5}}I^1_2s(1)s(4) \nonumber \\
&-\frac{2}{9\sqrt{5}}I^1_2s(1)s(5)+\frac{2}{9\sqrt{5}}I^1_2s(1)s(6)+\frac{1}{18\sqrt{5}}I^1_2s(2)s(3)
-\frac{1}{18\sqrt{5}}I^1_2s(2)s(4)-\frac{2}{9\sqrt{5}}I^1_2s(2)s(5)+\frac{2}{9\sqrt{5}}I^1_2s(2)s(6)
\nonumber \\
&-\frac{1}{18\sqrt{5}}I^1_2s(3)s(4)-\frac{2}{9\sqrt{5}}I^1_2s(3)s(5)+\frac{2}{9\sqrt{5}}I^1_2s(3)s(6)-\frac{1}{3\sqrt{5}}I^1_3s(1)s(2)-\frac{1}{3\sqrt{5}}I^1_3s(1)s(3)-\frac{1}{3\sqrt{5}}I^1_3s(2)s(3)], \nonumber
\end{align}
\begin{align}
\vert F^{27}_3 \rangle=&-\frac{\sqrt{10}}{36}I^1_1s(1)s(2)+\frac{\sqrt{10}}{72}I^1_1s(1)s(3)-\frac{\sqrt{10}}{72}I^1_1s(1)s(4)-\frac{\sqrt{10}}{72}I^1_1s(1)s(5)-\frac{\sqrt{10}}{360}I^1_1s(1)s(6)+\frac{\sqrt{10}}{72}I^1_1s(2)s(3)\nonumber \\
&-\frac{\sqrt{10}}{72}I^1_1s(2)s(4)-\frac{\sqrt{10}}{72}I^1_1s(2)s(5)-\frac{\sqrt{10}}{360}I^1_1s(2)s(6)+\frac{\sqrt{10}}{36}I^1_1s(3)s(4)+\frac{\sqrt{10}}{36}I^1_1s(3)s(5)+\frac{\sqrt{10}}{180}I^1_1s(3)s(6)\nonumber \\
&+\frac{\sqrt{10}}{90}I^1_2s(1)s(2)-\frac{\sqrt{10}}{180}I^1_2s(1)s(3)
+\frac{\sqrt{10}}{180}I^1_2s(1)s(4)+\frac{\sqrt{10}}{180}I^1_2s(1)s(5)-\frac{\sqrt{10}}{180}I^1_2s(1)s(6)-\frac{\sqrt{10}}{180}I^1_2s(2)s(3) \nonumber \\
&+\frac{\sqrt{10}}{180}I^1_2s(2)s(4)+\frac{\sqrt{10}}{180}I^1_2s(2)s(5)-\frac{\sqrt{10}}{180}I^1_2s(2)s(6)-\frac{\sqrt{10}}{90}I^1_2s(3)s(4)-\frac{\sqrt{10}}{90}I^1_2s(3)s(5)+\frac{\sqrt{10}}{90}I^1_2s(3)s(6)
\nonumber \\
&-\frac{\sqrt{10}}{30}I^1_2s(4)s(5)+\frac{\sqrt{10}}{30}I^1_2s(4)s(6)+\frac{\sqrt{10}}{30}I^1_2s(5)s(6)+\frac{\sqrt{10}}{30}I^1_3s(1)s(2)-\frac{\sqrt{10}}{60}I^1_3s(1)s(3)-\frac{\sqrt{10}}{20}I^1_3s(1)s(4) \nonumber \\
&-\frac{\sqrt{10}}{20}I^1_3s(1)s(5)+\frac{\sqrt{10}}{20}I^1_3s(1)s(6)-\frac{\sqrt{10}}{60}I^1_3s(2)s(3)-\frac{\sqrt{10}}{20}I^1_3s(2)s(4)-\frac{\sqrt{10}}{20}I^1_3s(2)s(5)+\frac{\sqrt{10}}{20}I^1_3s(2)s(6)], \nonumber
\end{align}
\begin{align}
\vert F^{27}_4 \rangle=&-\frac{\sqrt{30}}{72}I^1_1s(1)s(3)-\frac{\sqrt{30}}{72}I^1_1s(1)s(4)-\frac{\sqrt{30}}{72}I^1_1s(1)s(5)-\frac{\sqrt{30}}{360}I^1_1s(1)s(6)+\frac{\sqrt{30}}{72}I^1_1s(2)s(3)+\frac{\sqrt{30}}{72}I^1_1s(2)s(4)\nonumber \\
&+\frac{\sqrt{30}}{72}I^1_1s(2)s(5)+\frac{\sqrt{30}}{360}I^1_1s(2)s(6)+\frac{\sqrt{30}}{180}I^1_2s(1)s(3)+\frac{\sqrt{30}}{180}I^1_2s(1)s(4)+\frac{\sqrt{30}}{180}I^1_2s(1)s(5)-\frac{\sqrt{30}}{180}I^1_2s(1)s(6)\nonumber \\
&-\frac{\sqrt{30}}{180}I^1_2s(2)s(3)-\frac{\sqrt{30}}{180}I^1_2s(2)s(4)
-\frac{\sqrt{30}}{180}I^1_2s(2)s(5)+\frac{\sqrt{30}}{180}I^1_2s(2)s(6)+\frac{\sqrt{30}}{60}I^1_3s(1)s(3)+\frac{\sqrt{30}}{60}I^1_3s(1)s(4) \nonumber \\
&+\frac{\sqrt{30}}{60}I^1_3s(1)s(5)-\frac{\sqrt{30}}{60}I^1_3s(1)s(6)-\frac{\sqrt{30}}{60}I^1_3s(2)s(3)-\frac{\sqrt{30}}{60}I^1_3s(2)s(4)-\frac{\sqrt{30}}{60}I^1_3s(2)s(5)+\frac{\sqrt{30}}{60}I^1_3s(2)s(6)
\nonumber \\
&-\frac{1}{\sqrt{30}}I^1_3s(3)s(4)-\frac{1}{\sqrt{30}}I^1_3s(3)s(5)+\frac{1}{\sqrt{30}}I^1_3s(3)s(6)-\frac{1}{\sqrt{30}}I^1_3s(4)s(5)+\frac{1}{\sqrt{30}}I^1_3s(4)s(6)+\frac{1}{\sqrt{30}}I^1_3s(5)s(6)], \nonumber 
\end{align}
\begin{align}
\vert F^{27}_5 \rangle=&-\frac{\sqrt{10}}{180}I^1_1s(1)s(2)-\frac{\sqrt{10}}{180}I^1_1s(1)s(3)+\frac{\sqrt{10}}{180}I^1_1s(1)s(4)+\frac{\sqrt{10}}{180}I^1_1s(1)s(5)+\frac{\sqrt{10}}{90}I^1_1s(1)s(6)-\frac{\sqrt{10}}{180}I^1_1s(2)s(3)\nonumber \\
&+\frac{\sqrt{10}}{180}I^1_1s(2)s(4)+\frac{\sqrt{10}}{180}I^1_1s(2)s(5)+\frac{\sqrt{10}}{90}I^1_1s(2)s(6)+\frac{\sqrt{10}}{180}I^1_1s(3)s(4)+\frac{\sqrt{10}}{180}I^1_1s(3)s(5)+\frac{\sqrt{10}}{90}I^1_1s(3)s(6)\nonumber \\
&-\frac{\sqrt{10}}{60}I^1_1s(4)s(5)-\frac{\sqrt{10}}{30}I^1_1s(4)s(6)
-\frac{\sqrt{10}}{30}I^1_1s(5)s(6)+\frac{\sqrt{10}}{45}I^1_2s(1)s(2)+\frac{\sqrt{10}}{45}I^1_2s(1)s(3)-\frac{\sqrt{10}}{45}I^1_2s(1)s(4) \nonumber \\
&-\frac{\sqrt{10}}{45}I^1_2s(1)s(5)+\frac{\sqrt{10}}{45}I^1_2s(1)s(6)+\frac{\sqrt{10}}{45}I^1_2s(2)s(3)-\frac{\sqrt{10}}{45}I^1_2s(2)s(4)-\frac{\sqrt{10}}{45}I^1_2s(2)s(5)+\frac{\sqrt{10}}{45}I^1_2s(2)s(6)
\nonumber \\
&-\frac{\sqrt{10}}{45}I^1_2s(3)s(4)-\frac{\sqrt{10}}{45}I^1_2s(3)s(5)+\frac{\sqrt{10}}{45}I^1_2s(3)s(6)+\frac{\sqrt{10}}{15}I^1_3s(1)s(2)+\frac{\sqrt{10}}{15}I^1_3s(1)s(3)+\frac{\sqrt{10}}{15}I^1_3s(2)s(3)], \nonumber 
\end{align}
\begin{align}
\vert F^{27}_6 \rangle=&\frac{\sqrt{5}}{90}I^1_1s(1)s(2)-\frac{\sqrt{5}}{180}I^1_1s(1)s(3)+\frac{\sqrt{5}}{180}I^1_1s(1)s(4)-\frac{\sqrt{5}}{90}I^1_1s(1)s(5)-\frac{\sqrt{5}}{45}I^1_1s(1)s(6)-\frac{\sqrt{5}}{180}I^1_1s(2)s(3)+\nonumber \\
&\frac{\sqrt{5}}{180}I^1_1s(2)s(4)-\frac{\sqrt{5}}{90}I^1_1s(2)s(5)-\frac{\sqrt{5}}{45}I^1_1s(2)s(6)-\frac{\sqrt{5}}{90}I^1_1s(3)s(4)+\frac{\sqrt{5}}{45}I^1_1s(3)s(5)+\frac{2\sqrt{5}}{45}I^1_1s(3)s(6)-\nonumber \\
&\frac{2\sqrt{5}}{45}I^1_2s(1)s(2)+\frac{\sqrt{5}}{45}I^1_2s(1)s(3)
-\frac{\sqrt{5}}{45}I^1_2s(1)s(4)-\frac{\sqrt{5}}{180}I^1_2s(1)s(5)+\frac{\sqrt{5}}{180}I^1_2s(1)s(6)+\frac{\sqrt{5}}{45}I^1_2s(2)s(3)- \nonumber \\
&\frac{\sqrt{5}}{45}I^1_2s(2)s(4)-\frac{\sqrt{5}}{180}I^1_2s(2)s(5)+\frac{\sqrt{5}}{180}I^1_2s(2)s(6)+\frac{2\sqrt{5}}{45}I^1_2s(3)s(4)+\frac{\sqrt{5}}{90}I^1_2s(3)s(5)-\frac{\sqrt{5}}{90}I^1_2s(3)s(6)+
\nonumber \\
&\frac{\sqrt{5}}{30}I^1_2s(4)s(5)-\frac{\sqrt{5}}{30}I^1_2s(4)s(6)+\frac{\sqrt{5}}{15}I^1_2s(5)s(6)+\frac{\sqrt{5}}{15}I^1_3s(1)s(2)-\frac{\sqrt{5}}{30}I^1_3s(1)s(3)-\frac{\sqrt{5}}{10}I^1_3s(1)s(4)+
\nonumber \\
&\frac{\sqrt{5}}{20}I^1_3s(1)s(5)-\frac{\sqrt{5}}{20}I^1_3s(1)s(6)-\frac{\sqrt{5}}{30}I^1_3s(2)s(3)-\frac{\sqrt{5}}{10}I^1_3s(2)s(4)+\frac{\sqrt{5}}{20}I^1_3s(2)s(5)-\frac{\sqrt{5}}{20}I^1_3s(2)s(6)], \nonumber 
\end{align}
\begin{align}
\vert F^{27}_7 \rangle=&\frac{\sqrt{15}}{180}I^1_1s(1)s(3)+\frac{\sqrt{15}}{180}I^1_1s(1)s(4)-\frac{\sqrt{15}}{90}I^1_1s(1)s(5)-\frac{\sqrt{15}}{45}I^1_1s(1)s(6)-\frac{\sqrt{15}}{180}I^1_1s(2)s(3)-\frac{\sqrt{15}}{180}I^1_1s(2)s(4)+\nonumber \\
&\frac{\sqrt{15}}{90}I^1_1s(2)s(5)+\frac{\sqrt{15}}{45}I^1_1s(2)s(6)-\frac{\sqrt{15}}{45}I^1_2s(1)s(3)-\frac{\sqrt{15}}{45}I^1_2s(1)s(4)-\frac{\sqrt{15}}{180}I^1_2s(1)s(5)+\frac{\sqrt{15}}{180}I^1_2s(1)s(6)+\nonumber \\
&\frac{\sqrt{15}}{45}I^1_2s(2)s(3)+\frac{\sqrt{15}}{45}I^1_2s(2)s(4)
+\frac{\sqrt{15}}{180}I^1_2s(2)s(5)-\frac{\sqrt{15}}{180}I^1_2s(2)s(6)+\frac{\sqrt{15}}{30}I^1_3s(1)s(3)+\frac{\sqrt{15}}{30}I^1_3s(1)s(4)- \nonumber \\
&\frac{\sqrt{15}}{60}I^1_3s(1)s(5)+\frac{\sqrt{15}}{60}I^1_3s(1)s(6)-\frac{\sqrt{15}}{30}I^1_3s(2)s(3)-\frac{\sqrt{15}}{30}I^1_3s(2)s(4)+\frac{\sqrt{15}}{60}I^1_3s(2)s(5)-\frac{\sqrt{15}}{60}I^1_3s(2)s(6)-
\nonumber \\
&\frac{1}{\sqrt{15}}I^1_3s(3)s(4)+\frac{1}{2\sqrt{15}}I^1_3s(3)s(5)-\frac{1}{2\sqrt{15}}I^1_3s(3)s(6)+\frac{1}{2\sqrt{15}}I^1_3s(4)s(5)-\frac{1}{2\sqrt{15}}I^1_3s(4)s(6)+\frac{1}{\sqrt{15}}I^1_3s(5)s(6)], \nonumber 
\end{align}
\begin{align}
\vert F^{27}_8 \rangle=&-\frac{\sqrt{15}}{90}I^1_1s(1)s(2)+\frac{\sqrt{15}}{180}I^1_1s(1)s(3)+\frac{\sqrt{15}}{180}I^1_1s(1)s(4)+\frac{\sqrt{15}}{180}I^1_1s(2)s(3)+\frac{\sqrt{15}}{180}I^1_1s(2)s(4)-\frac{\sqrt{15}}{90}I^1_1s(3)s(4) \nonumber \\
&-\frac{\sqrt{15}}{45}I^1_2s(1)s(2)+\frac{\sqrt{15}}{90}I^1_2s(1)s(3)+\frac{\sqrt{15}}{60}I^1_2s(1)s(4)+\frac{\sqrt{15}}{60}I^1_2s(1)s(5)+\frac{\sqrt{15}}{60}I^1_2s(1)s(6)+\frac{\sqrt{15}}{90}I^1_2s(2)s(3) \nonumber \\
&+\frac{\sqrt{15}}{90}I^1_2s(2)s(4)+\frac{\sqrt{15}}{60}I^1_2s(2)s(5)
+\frac{\sqrt{15}}{60}I^1_2s(2)s(6)-\frac{\sqrt{15}}{45}I^1_2s(3)s(4)-\frac{\sqrt{15}}{30}I^1_2s(3)s(5)-\frac{\sqrt{15}}{30}I^1_2s(3)s(6)  \nonumber \\
&-\frac{\sqrt{15}}{30}I^1_2s(4)s(5)-\frac{\sqrt{15}}{30}I^1_2s(4)s(6)+\frac{\sqrt{15}}{20}I^1_3s(1)s(5)+\frac{\sqrt{15}}{20}I^1_3s(1)s(6)+\frac{\sqrt{15}}{20}I^1_3s(2)s(5)+\frac{\sqrt{15}}{20}I^1_3s(2)s(6)], \nonumber 
\end{align}
\begin{align}
\vert F^{27}_9 \rangle=&\frac{\sqrt{5}}{60}I^1_1s(1)s(3)-\frac{\sqrt{5}}{60}I^1_1s(1)s(4)-\frac{\sqrt{5}}{60}I^1_1s(2)s(3)+\frac{\sqrt{5}}{60}I^1_1s(2)s(4)+\frac{\sqrt{5}}{30}I^1_2s(1)s(3)-\frac{\sqrt{5}}{30}I^1_2s(1)s(4)- \nonumber \\
&\frac{\sqrt{5}}{20}I^1_2s(1)s(5)-\frac{\sqrt{5}}{20}I^1_2s(1)s(6)-\frac{\sqrt{5}}{30}I^1_2s(2)s(3)+\frac{\sqrt{5}}{30}I^1_2s(2)s(4)+\frac{\sqrt{5}}{20}I^1_2s(2)s(5)+\frac{\sqrt{5}}{20}I^1_2s(2)s(6)+ \nonumber \\
&\frac{\sqrt{5}}{20}I^1_3s(1)s(5)+\frac{\sqrt{5}}{20}I^1_3s(1)s(6)
-\frac{\sqrt{5}}{20}I^1_3s(2)s(5)-\frac{\sqrt{5}}{20}I^1_3s(2)s(6)-\frac{\sqrt{5}}{10}I^1_3s(3)s(5)-\frac{\sqrt{5}}{10}I^1_3s(3)s(6)+  \nonumber \\
&\frac{\sqrt{5}}{10}I^1_3s(4)s(5)+\frac{\sqrt{5}}{10}I^1_3s(4)s(6)].
\end{align}
$F^{27}$ multiplet for I=2; \\
\begin{align}
\vert F^{27}_1 \rangle=\frac{1}{4\sqrt{15}}[&12I^2s(5)s(6)-3I^2s(4)s(6)-3I^2s(3)s(6)-3I^2s(2)s(6)-3I^2s(1)s(6)-3I^2s(4)s(5)
-3I^2s(3)s(5)-\nonumber \\
&3I^2s(2)s(5)-3I^2s(1)s(5)+2I^2s(3)s(4)+2I^2s(2)s(4)+2I^2s(1)s(4)+2I^2s(2)s(3)+2I^2s(1)s(3)+
\nonumber \\
&2I^2s(1)s(2)], \nonumber
\end{align}
\begin{align}
\vert F^{27}_2 \rangle=\frac{1}{12}[&2I^2s(1)s(2)+2I^2s(1)s(3)-2I^2s(1)s(4)+I^2s(1)s(5)-3I^2s(1)s(6)+2I^2s(2)s(3)
-2I^2s(2)s(4)+\nonumber \\
&I^2s(2)s(5)-3I^2s(2)s(6)-2I^2s(3)s(4)+I^2s(3)s(5)-3I^2s(3)s(6)-3I^2s(4)s(5)+9I^2s(4)s(6)],
\nonumber 
\end{align}
\begin{align}
\vert F^{27}_3 \rangle=\frac{1}{6\sqrt{2}}[&2I^2s(1)s(2)-I^2s(1)s(3)+I^2s(1)s(4)+I^2s(1)s(5)-3I^2s(1)s(6)-I^2s(2)s(3)
+I^2s(2)s(4)+\nonumber \\
&I^2s(2)s(5)-3I^2s(2)s(6)-2I^2s(3)s(4)-2I^2s(3)s(5)+6I^2s(3)s(6)],
\nonumber 
\end{align}
\begin{align}
\vert F^{27}_4 \rangle=\frac{1}{2\sqrt{6}}[&I^2s(1)s(3)+I^2s(1)s(4)+I^2s(1)s(5)-3I^2s(1)s(6)-I^2s(2)s(3)-I^2s(2)s(4)
-I^2s(2)s(5)+\nonumber \\
&3I^2s(2)s(6)],
\nonumber 
\end{align}
\begin{align}
\vert F^{27}_5 \rangle=\frac{1}{3\sqrt{2}}[&I^2s(1)s(2)+I^2s(1)s(3)-I^2s(1)s(4)-I^2s(1)s(5)+I^2s(2)s(3)-I^2s(2)s(4)-I^2s(2)s(5)-\nonumber \\
&I^2s(3)s(4)-I^2s(3)s(5)+3I^2s(4)s(5)],
\nonumber 
\end{align}
\begin{align}
\vert F^{27}_6 \rangle=\frac{1}{6\sqrt{2}}[&2I^2s(1)s(2)-I^2s(1)s(3)+I^2s(1)s(4)-2I^2s(1)s(5)-I^2s(2)s(3)+I^2s(2)s(4)-2I^2s(2)s(5)-\nonumber \\
&2I^2s(3)s(4)+4I^2s(3)s(5)],
\nonumber 
\end{align}
\begin{align}
\vert F^{27}_7 \rangle=\frac{1}{2\sqrt{3}}[&I^2s(1)s(3)+I^2s(1)s(4)-2I^2s(1)s(5)-I^2s(2)s(3)-I^2s(2)s(4)+2I^2s(2)s(5)],\nonumber 
\end{align}
\begin{align}
\vert F^{27}_8 \rangle=\frac{1}{2\sqrt{3}}[&2I^2s(1)s(2)-I^2s(1)s(3)-I^2s(1)s(4)-I^2s(2)s(3)-I^2s(2)s(4)+2I^2s(3)s(4)],\nonumber 
\end{align}
\begin{align}
\vert F^{27}_9 \rangle=\frac{1}{2}[&I^2s(1)s(3)-I^2s(1)s(4)-I^2s(2)s(3)+I^2s(2)s(4)]. 
\end{align}
$F^{28}$ multiplet for I=2; \\
\begin{align}
\vert F^{28} \rangle=\frac{1}{\sqrt{15}}[&I^2s(1)s(2)+I^2s(1)s(3)+I^2s(1)s(4)+I^2s(1)s(5)+I^2s(1)s(6)+I^2s(2)s(3)
+I^2s(2)s(4)+\nonumber \\
&I^2s(2)s(5)+I^2s(2)s(6)+I^2s(3)s(4)+I^2s(3)s(5)+I^2s(3)s(6)+I^2s(4)s(5)+I^2s(4)s(6)+
\nonumber \\
&I^2s(5)s(6)].
\end{align}
$F^{1}$ multiplet for I=0; \\
\begin{align}
\vert F^{1}_1 \rangle=\frac{1}{8\sqrt{18}}[&4I^0_1s(5)s(6)-2I^0_1s(4)s(6)-2I^0_1s(4)s(5)+3I^0_2s(2)s(5)-2I^0_1s(2)s(3)+4I^0_1s(1)s(2)
-2I^0_1s(1)s(3)+\nonumber \\
&3I^0_2s(2)s(6)+3I^0_2s(1)s(5)-2I^0_1s(3)s(6)+I^0_1s(2)s(6)+4I^0_1s(3)s(4)-2I^0_1s(3)s(5)-2I^0_1s(2)s(4)+
\nonumber \\
&I^0_1s(2)s(5)+I^0_1s(1)s(6)+3I^0_2s(1)s(6)+I^0_1s(1)s(5)-2I^0_1s(1)s(4)], \nonumber
\end{align}
\begin{align}
\vert F^{1}_2 \rangle=\frac{\sqrt{2}}{16\sqrt{3}}[&4I^0_2s(5)s(6)-2I^0_2s(4)s(6)-2I^0_2s(4)s(5)+2I^0_2s(3)s(5)+I^0_2s(2)s(5)-2I^0_1s(1)s(3)
+2I^0_1s(1)s(3)+\nonumber \\
&2I^0_2s(3)s(6)+I^0_2s(2)s(6)-I^0_2s(1)s(5)-I^0_1s(2)s(6)+2I^0_1s(2)s(4)-I^0_1s(2)s(5)+I^0_1s(1)s(6)-
\nonumber \\
&I^0_2s(1)s(6)+I^0_1s(1)s(5)-2I^0_1s(1)s(4)], \nonumber
\end{align}
\begin{align}
\vert F^{1}_3 \rangle=\frac{1}{8\sqrt{6}}[&2I^0_1s(4)s(6)-2I^0_1s(4)s(5)+2I^0_2s(2)s(4)-2I^0_2s(2)s(3)+4I^0_2s(1)s(2)-2I^0_2s(1)s(3)
-I^0_2s(2)s(6)+\nonumber \\
&I^0_2s(2)s(5)+2I^0_2s(1)s(4)-2I^0_1s(3)s(6)+I^0_1s(2)s(6)+2I^0_1s(3)s(5)-I^0_1s(2)s(5)+I^0_1s(1)s(6)-
\nonumber \\
&I^0_2s(1)s(6)-I^0_1s(1)s(5)+I^0_2s(1)s(5)], \nonumber
\end{align}
\begin{align}
\vert F^{1}_4 \rangle=\frac{1}{24\sqrt{2}}[&6I^0_2s(4)s(6)-6I^0_2s(4)s(5)+4I^0_2s(3)s(4)-6I^0_2s(2)s(3)+6I^0_2s(1)s(3)-2I^0_2s(3)s(6)
+2I^0_2s(3)s(5)+\nonumber \\
&2I^0_2s(2)s(4)-2I^0_2s(1)s(4)-3I^0_1s(2)s(6)-I^0_2s(2)s(6)+3I^0_1s(2)s(5)+I^0_2s(2)s(5)+3I^0_1s(1)s(6)+
\nonumber \\
&I^0_2s(1)s(6)-3I^0_1s(1)s(5)-I^0_2s(1)s(5)], \nonumber
\end{align}
\begin{align}
\vert F^{1}_5 \rangle=\frac{1}{6}[&I^0_2s(3)s(6)-I^0_2s(3)s(5)+I^0_2s(3)s(4)-I^0_2s(2)s(6)+I^0_2s(2)s(5)-I^0_2s(2)s(4)
+I^0_2s(1)s(6)-\nonumber \\
&I^0_2s(1)s(5)+I^0_2s(1)s(4)].
\end{align}

\end{widetext}
We note that the flavor multiplet bases are orthonormal to each other, that is, $\langle F^i_k \vert F^j_l
\rangle$ = ${\delta}_{ij}$${\delta}_{kl}$.

\section{CS coupling}

In this Appendix B, we present the color $\otimes$ spin basis, which is obtained from the CS coupling scheme. As mentioned in subsection B of section III. the CG coefficient of combining the color singlet basis with the S=0 basis is calcutated by using Eq.~(\ref{eq-CG}). The color $\otimes$ spin basis represented by Young-Yamanouci basis [2,2,1,1] and [3,3] is given as follows;
\begin{widetext}\allowdisplaybreaks
Young-Yamanouci basis [2,2,1,1]; \\
\begin{align}
\vert [C,S^0]_1 \rangle=&-\frac{\sqrt{6}}{4}\vert C_1 \rangle \otimes \vert S^0_4 \rangle+
\frac{\sqrt{6}}{4}\vert C_2 \rangle \otimes \vert S^0_5 \rangle+
\frac{\sqrt{6}}{12}\vert C_3 \rangle \otimes \vert S^0_2 \rangle-
\frac{\sqrt{3}}{6}\vert C_3 \rangle \otimes \vert S^0_1 \rangle-
\frac{\sqrt{6}}{12}\vert C_4 \rangle \otimes \vert S^0_3 \rangle
\nonumber \\
&-\frac{\sqrt{3}}{6}\vert C_5 \rangle \otimes \vert S^0_3 \rangle, \nonumber
\end{align}
\begin{align}
\vert [C,S^0]_2 \rangle=&-\frac{\sqrt{3}}{6}\vert C_4 \rangle \otimes \vert S^0_1 \rangle-
\frac{\sqrt{6}}{12}\vert C_4 \rangle \otimes \vert S^0_2 \rangle+
\frac{\sqrt{3}}{6}\vert C_5 \rangle \otimes \vert S^0_2 \rangle-
\frac{\sqrt{6}}{12}\vert C_3 \rangle \otimes \vert S^0_3 \rangle+
\frac{\sqrt{6}}{4}\vert C_2 \rangle \otimes \vert S^0_4 \rangle 
\nonumber \\
&+\frac{\sqrt{6}}{4}\vert C_1 \rangle \otimes \vert S^0_5 \rangle, \nonumber
\end{align}
\begin{align}
\vert [C,S^0]_3 \rangle=&-\frac{\sqrt{3}}{6}\vert C_1 \rangle \otimes \vert S^0_1 \rangle-
\frac{1}{3}\vert C_3 \rangle \otimes \vert S^0_1 \rangle+
\frac{\sqrt{6}}{12}\vert C_1 \rangle \otimes \vert S^0_2 \rangle-
\frac{\sqrt{2}}{3}\vert C_3 \rangle \otimes \vert S^0_2 \rangle-
\frac{\sqrt{6}}{12}\vert C_2 \rangle \otimes \vert S^0_3 \rangle 
\nonumber \\
&+\frac{\sqrt{2}}{3}\vert C_4 \rangle \otimes \vert S^0_3 \rangle-
\frac{1}{3}\vert C_5 \rangle \otimes \vert S^0_3 \rangle-
\frac{\sqrt{6}}{12}\vert C_3 \rangle \otimes \vert S^0_4 \rangle+
\frac{\sqrt{6}}{12}\vert C_4 \rangle \otimes \vert S^0_5 \rangle+
\frac{\sqrt{3}}{6}\vert C_5 \rangle \otimes \vert S^0_5 \rangle, \nonumber
\end{align}
\begin{align}
\vert [C,S^0]_4 \rangle=&-\frac{\sqrt{3}}{6}\vert C_2 \rangle \otimes \vert S^0_1 \rangle-
\frac{1}{3}\vert C_4 \rangle \otimes \vert S^0_1 \rangle-
\frac{\sqrt{6}}{12}\vert C_2 \rangle \otimes \vert S^0_2 \rangle+
\frac{\sqrt{2}}{3}\vert C_4 \rangle \otimes \vert S^0_2 \rangle+
\frac{1}{3}\vert C_5 \rangle \otimes \vert S^0_2 \rangle 
\nonumber \\
&-\frac{\sqrt{6}}{12}\vert C_1 \rangle \otimes \vert S^0_3 \rangle+
\frac{\sqrt{2}}{3}\vert C_3 \rangle \otimes \vert S^0_3 \rangle+
\frac{\sqrt{6}}{12}\vert C_4 \rangle \otimes \vert S^0_4 \rangle-
\frac{\sqrt{3}}{6}\vert C_5 \rangle \otimes \vert S^0_4 \rangle+
\frac{\sqrt{6}}{12}\vert C_3 \rangle \otimes \vert S^0_5 \rangle, \nonumber
\end{align}
\begin{align}
\vert [C,S^0]_5 \rangle=&\frac{2}{3}\vert C_5 \rangle \otimes \vert S^0_1 \rangle+
\frac{\sqrt{3}}{6}\vert C_2 \rangle \otimes \vert S^0_2 \rangle+
\frac{1}{3}\vert C_4 \rangle \otimes \vert S^0_2 \rangle-
\frac{\sqrt{3}}{6}\vert C_1 \rangle \otimes \vert S^0_3 \rangle-
\frac{1}{3}\vert C_3 \rangle \otimes \vert S^0_3 \rangle- 
\nonumber \\
&\frac{\sqrt{3}}{6}\vert C_4 \rangle \otimes \vert S^0_4 \rangle+
\frac{\sqrt{3}}{6}\vert C_3 \rangle \otimes \vert S^0_5 \rangle, \nonumber
\end{align}
\begin{align}
\vert [C,S^0]_6 \rangle=&-\frac{\sqrt{6}}{6}\vert C_1 \rangle \otimes \vert S^0_1 \rangle+
\frac{\sqrt{2}}{6}\vert C_3 \rangle \otimes \vert S^0_1 \rangle+
\frac{\sqrt{3}}{6}\vert C_1 \rangle \otimes \vert S^0_2 \rangle+
\frac{1}{3}\vert C_3 \rangle \otimes \vert S^0_2 \rangle-
\frac{\sqrt{3}}{6}\vert C_2 \rangle \otimes \vert S^0_3 \rangle 
\nonumber \\
&-\frac{1}{3}\vert C_4 \rangle \otimes \vert S^0_3 \rangle+
\frac{\sqrt{2}}{6}\vert C_5 \rangle \otimes \vert S^0_3 \rangle-
\frac{\sqrt{3}}{6}\vert C_3 \rangle \otimes \vert S^0_4 \rangle+
\frac{\sqrt{3}}{6}\vert C_4 \rangle \otimes \vert S^0_5 \rangle+
\frac{\sqrt{6}}{6}\vert C_5 \rangle \otimes \vert S^0_5 \rangle, \nonumber
\end{align}
\begin{align}
\vert [C,S^0]_7 \rangle=&-\frac{\sqrt{6}}{6}\vert C_2 \rangle \otimes \vert S^0_1 \rangle+
\frac{\sqrt{2}}{6}\vert C_4 \rangle \otimes \vert S^0_1 \rangle-
\frac{\sqrt{3}}{6}\vert C_2 \rangle \otimes \vert S^0_2 \rangle-
\frac{1}{3}\vert C_4 \rangle \otimes \vert S^0_2 \rangle-
\frac{\sqrt{2}}{6}\vert C_5 \rangle \otimes \vert S^0_2 \rangle 
\nonumber \\
&-\frac{\sqrt{3}}{6}\vert C_1 \rangle \otimes \vert S^0_3 \rangle-
\frac{1}{3}\vert C_3 \rangle \otimes \vert S^0_3 \rangle+
\frac{\sqrt{3}}{6}\vert C_4 \rangle \otimes \vert S^0_4 \rangle-
\frac{\sqrt{6}}{6}\vert C_5 \rangle \otimes \vert S^0_4 \rangle+
\frac{\sqrt{3}}{6}\vert C_3 \rangle \otimes \vert S^0_5 \rangle, \nonumber
\end{align}
\begin{align}
\vert [C,S^0]_8 \rangle=&-\frac{\sqrt{2}}{3}\vert C_5 \rangle \otimes \vert S^0_1 \rangle+
\frac{\sqrt{6}}{6}\vert C_2 \rangle \otimes \vert S^0_2 \rangle-
\frac{\sqrt{2}}{6}\vert C_4 \rangle \otimes \vert S^0_2 \rangle-
\frac{\sqrt{6}}{6}\vert C_1 \rangle \otimes \vert S^0_3 \rangle+
\frac{\sqrt{2}}{6}\vert C_3 \rangle \otimes \vert S^0_3 \rangle 
\nonumber \\
&-\frac{\sqrt{6}}{6}\vert C_4 \rangle \otimes \vert S^0_4 \rangle+
\frac{\sqrt{6}}{6}\vert C_3 \rangle \otimes \vert S^0_5 \rangle,
 \nonumber
\end{align}
\begin{align}
\vert [C,S^0]_9 \rangle=&\frac{\sqrt{2}}{\sqrt{15}}\vert C_5 \rangle \otimes \vert S^0_1 \rangle-
\frac{\sqrt{2}}{\sqrt{15}}\vert C_4 \rangle \otimes \vert S^0_2 \rangle+
\frac{\sqrt{2}}{\sqrt{15}}\vert C_3 \rangle \otimes \vert S^0_3 \rangle-
\frac{\sqrt{3}}{\sqrt{10}}\vert C_2 \rangle \otimes \vert S^0_4 \rangle+
\frac{\sqrt{3}}{\sqrt{10}}\vert C_1 \rangle \otimes \vert S^0_5 \rangle.
\end{align}
Young-Yamanouci basis [3,3]; \\
\begin{align}
\vert [C,S^0]_1 \rangle=&\frac{1}{2}[\vert C_1 \rangle \otimes \vert S^0_2 \rangle+
\vert C_2 \rangle \otimes \vert S^0_3 \rangle+
\vert C_3 \rangle \otimes \vert S^0_4 \rangle+
\vert C_4 \rangle \otimes \vert S^0_5 \rangle], \nonumber
\end{align}
\begin{align}
\vert [C,S^0]_2 \rangle=&\frac{1}{2}\vert C_1 \rangle \otimes \vert S^0_1 \rangle+
\frac{\sqrt{2}}{4}\vert C_1 \rangle \otimes \vert S^0_2 \rangle-
\frac{\sqrt{2}}{4}\vert C_2 \rangle \otimes \vert S^0_3 \rangle-
\frac{\sqrt{2}}{4}\vert C_3 \rangle \otimes \vert S^0_4 \rangle+
\frac{\sqrt{2}}{4}\vert C_4 \rangle \otimes \vert S^0_5 \rangle- \nonumber \\
&\frac{1}{2}\vert C_5 \rangle \otimes \vert S^0_5 \rangle, \nonumber
\end{align}
\begin{align}
\vert [C,S^0]_3 \rangle=&-\frac{\sqrt{2}}{4}\vert C_1 \rangle \otimes \vert S^0_3 \rangle+
\frac{1}{2}\vert C_2 \rangle \otimes \vert S^0_1 \rangle-
\frac{\sqrt{2}}{4}\vert C_2 \rangle \otimes \vert S^0_2 \rangle+
\frac{\sqrt{2}}{4}\vert C_3 \rangle \otimes \vert S^0_5 \rangle+
\frac{\sqrt{2}}{4}\vert C_4 \rangle \otimes \vert S^0_4 \rangle \nonumber \\
&+\frac{1}{2}\vert C_5 \rangle \otimes \vert S^0_4 \rangle, \nonumber
\end{align}
\begin{align}
\vert [C,S^0]_4 \rangle=&-\frac{\sqrt{2}}{4}\vert C_1 \rangle \otimes \vert S^0_4 \rangle+
\frac{\sqrt{2}}{4}\vert C_2 \rangle \otimes \vert S^0_5 \rangle+
\frac{1}{2}\vert C_3 \rangle \otimes \vert S^0_1 \rangle-
\frac{\sqrt{2}}{4}\vert C_3 \rangle \otimes \vert S^0_2 \rangle+
\frac{\sqrt{2}}{4}\vert C_4 \rangle \otimes \vert S^0_3 \rangle \nonumber \\
&+\frac{1}{2}\vert C_5 \rangle \otimes \vert S^0_3 \rangle, \nonumber
\end{align}
\begin{align}
\vert [C,S^0]_5 \rangle=&\frac{\sqrt{2}}{4}\vert C_1 \rangle \otimes \vert S^0_5 \rangle+
\frac{\sqrt{2}}{4}\vert C_2 \rangle \otimes \vert S^0_4 \rangle+
\frac{\sqrt{2}}{4}\vert C_3 \rangle \otimes \vert S^0_3 \rangle+
\frac{1}{2}\vert C_4 \rangle \otimes \vert S^0_1 \rangle+
\frac{\sqrt{2}}{4}\vert C_4 \rangle \otimes \vert S^0_2 \rangle- \nonumber \\
&\frac{1}{2}\vert C_5 \rangle \otimes \vert S^0_2 \rangle. 
\end{align}

\end{widetext}

\end{document}